\documentclass[fleqn,usenatbib]{mnras}

\usepackage{newtxtext,newtxmath}

\usepackage[T1]{fontenc}
\usepackage{ae,aecompl}

\usepackage{graphicx}	
\usepackage{amsmath}	
\usepackage{amssymb}	
\usepackage{ctable}
\usepackage{color}       

\title[Intervening HI absorption at $0.4<z<1.0$]{A successful search for intervening 21\,cm HI absorption in galaxies at $0.4<z<$1.0 with the Australian Square Kilometre Array Pathfinder (ASKAP) } 

\author[Elaine M. Sadler et al.]{Elaine M. Sadler$^{1,2,3}$\thanks{E-mail: elaine.sadler@sydney.edu.au}, 
Vanessa A. Moss$^{2,4}$, 
James R. Allison$^{5,3}$, 
Elizabeth K. Mahony$^{2}$, 
\newauthor
Matthew T. Whiting$^{2}$, Helen M. Johnston$^{1}$, Sara L. Ellison$^{6}$, Claudia del P. Lagos$^{7,3}$, 
\newauthor
B\"arbel S. Koribalski$^{2}$
\vspace*{6pt}
\\
$^1$Sydney Institute for Astronomy, School of Physics, University of Sydney, NSW 2006, Australia\\
$^2$ATNF, CSIRO Astronomy and Space Science,  PO Box 76, Epping, NSW 1710, Australia \\
$^3$ARC Centre of Excellence for All Sky Astrophysics in 3 Dimensions (ASTRO 3D) \\
$^4$ASTRON, The Netherlands Institute for Radio Astronomy, Postbus 2, NL-7900 AA Dwingeloo, the Netherlands \\
$^5$Sub-Dept. of Astrophysics, Department of Physics, University of Oxford, Denys Wilkinson Building, Keble Rd., Oxford, OX1 3RH, UK\\
$^6$Department of Physics \& Astronomy, University of Victoria, Finnerty Road, Victoria, British Columbia V8P 1A1, Canada \\
$^7$International Centre for Radio Astronomy Research, The University of Western Australia, Crawley, WA 6009, Australia \\
\\}

\date{Accepted XXX. Received YYY; in original form ZZZ}

\pubyear{2020}

\begin{document}
\label{firstpage}
\pagerange{\pageref{firstpage}--\pageref{lastpage}}
\maketitle

\begin{abstract}
We have used the Australian Square Kilometre Array Pathfinder (ASKAP) radio telescope to search for intervening 21\,cm neutral hydrogen (HI) absorption along the line of sight to 53 bright radio continuum sources. 
Our observations are sensitive to HI column densities typical of Damped Lyman Alpha absorbers (DLAs) in cool gas with an HI spin temperature below about 300--500\,K. 
The six-dish Boolardy Engineering Test Array (BETA) and twelve-antenna Early Science array (ASKAP-12) covered a frequency range corresponding to redshift $0.4<z<1.0$ and $0.37<z<0.77$ respectively for the HI line. Fifty of the 53 radio sources observed have reliable optical redshifts, giving a total redshift path $\Delta z$ = 21.37. This was a spectroscopically \-untargeted  survey, with no prior assumptions about the location of the lines in redshift space. Four intervening HI lines were detected, two of them new. 
In each case, the estimated HI column density lies above the DLA limit for HI spin temperatures above 50-80\,K, and we estimate a DLA number density at redshift $z\sim0.6$ of $n(z)=0.19\substack{+0.15 \\ -0.09}$. 
This value lies somewhat above the general trend of $n(z)$ with redshift seen in optical DLA studies.   
Although the current sample is small, it represents an important proof of concept for the much larger 21\,cm {\it First Large Absorption Survey in HI}\ (FLASH) project to be carried out with the full 36-antenna ASKAP telescope, probing a total redshift path $\Delta z\sim\,50,000$. 
\end{abstract}




\section{Introduction}
\subsection{Neutral hydrogen in the distant Universe} 
Almost all our current knowledge about the amount and distribution of neutral hydrogen (HI) beyond the local (redshift $z\lesssim0.2$) Universe comes from optical studies of QSO absorption lines. The strongest of these lines, the Damped Lyman-$\alpha$ Absorbers (DLAs), arise in sight-lines with HI column densities above $2\times10^{20}$ atoms cm$^{-2}$, i.e. similar to the typical column density in the Milky Way disk \citep[]{wolfe86,wolfe05}. DLAs are thought to account for the bulk of the cosmic HI mass density across a wide range in redshift, and so are important tracers of the neutral-gas reservoirs for star formation in the distant Universe. 

Large optical QSO surveys for DLAs are only possible at redshift $z\gtrsim1.7$, where the ultraviolet Ly$\alpha$ absorption line is redshifted to wavelengths visible with ground-based telescopes. At lower redshifts the Mg II absorption line has been used to select candidate DLA systems \citep[]{briggs83,rao00,rao06,ellison06,kanekar09,rao17}, but follow-up ultraviolet spectroscopy with the \textit{Hubble Space Telescope} (\textit{HST}) is then needed to measure the Ly$\alpha$ line profile and HI column density so the samples studied are much smaller. In addition, it is not straightforward to understand and quantify the selection biases present in these pre-selected samples and this can introduce further uncertainties in estimates of the cosmic HI mass density at $z<1.7$ \citep[][]{rao06,neeleman16,rao17,berg17}.  

Other approaches have been used to probe HI in the redshift range $0.2\lesssim z\lesssim1.7$, including HI emission-line stacking \citep[][]{lah07,delhaize13,kanekar16} and measurements of the UV spectra of QSOs observed with the Hubble Space Telescope without any MgII preselection  \citep{neeleman16}, but the amount and distribution of HI in galaxies in this redshift range remains poorly constrained. 

The extent to which optical DLA surveys are affected by dust obscuration remains unclear. Absorbers with a high HI column density may contain enough dust to redden and dim the light of background QSOs so that they are missed by colour-selected, flux-limited samples \citep{fall89}. A study of radio-selected QSOs by \cite{ellison05} suggests that the effect is generally small, with only a modest increase in reddening for typical absorbers (mean $E(B-V)<0.04$\,mag), though individual DLA systems with much higher reddening have also been identified \citep{heintz18,geier19}.  
\cite{pontzen09} estimate that around 7\,percent of DLAs at redshift $1.8<z<3.5$ are missing from optical samples due to dust obscuration, while \cite{krogager19} suggest that optical DLA studies underestimate the cosmic mass density of neutral hydrogen by 10--50\,percent at $z\sim3$, and by up to a factor of two at $z=2.2$. Previous studies have concentrated mainly on ground-based surveys for DLAs at $z>1.7$, and the effects of dust obscuration in lower-redshift DLA systems at $z<1$ remain to be quantified. 

In principle, radio measurements of the 21\,cm HI absorption line provide an alternative tool for identifying DLA systems, particularly at redshift $z<1.7$ where ground-based optical DLA studies are not possible \citep[e.g.][]{kanekar04,morganti15}. Importantly, because the optical depth of the 21-cm absorption line is inversely related to its excitation (spin) temperature, it is most sensitive to the cold neutral medium (CNM; $T \sim 100$\,K). This is the component of the HI most likely to trace star formation in galaxies and therefore directly follow its evolution throughout cosmic history (see e.g. \citealt{kanekar14a} and references therein). Since 21\,cm HI absorption measurements in the radio are unaffected by dust obscuration along the line of sight, they may also help clarify the extent to which optical DLA studies have been affected by a dust bias. 

A handful of 21\,cm DLAs have been detected in blind\footnote{By a `blind' search, we mean a search that is spectroscopically untargeted and uses no prior assumptions about the redshift of any HI absorption lines} searches with single-dish radio telescopes \citep[][]{brown73,brown83,darling04}, but these searches were limited both by the available spectral bandpass and by the effects of terrestrial radio interference 
\citep[e.g.][]{edel16}. As noted by \cite{wolfe05}, the redshift interval covered by a single optical spectrum, $\Delta z\sim1$, is large compared to that sampled by early radio 21\,cm surveys (typically $\Delta z\sim0.02$) so that optical surveys were more efficient at covering a large redshift path length.   
These earlier limitations have now largely been removed through (i) the advent of new wide-band correlators for radio interferometers that provide instantaneous redshift coverage approaching that of optical spectrographs, and (ii) the construction of new SKA pathfinder and precursor radio telescopes on radio-quiet sites.  

In this paper, we present the results from a blind 21\,cm search towards 53 bright radio sources for HI absorption in galaxies at redshift $0.4<z<1.0$. The observations were carried out in commissioning time with the first 6--12 dishes of the Australian SKA Pathfinder (ASKAP) telescope \citep{mcconnell16}. This work is intended to pave the way for a blind, all-sky 21\,cm absorption survey, the First Large Absorption Survey in HI (FLASH) to be carried out with the full 36-dish ASKAP telescope. Recent early science results from FLASH have discovered an intervening 21-cm absorber in the 50-square-degree GAMA\,23 field, demonstrating the feasibility of such a wide-field survey (\citealt{allison20}). However, FLASH has not yet covered enough sky area to return a high detection yield. By targeting a modest sample of very bright radio sources, we aim here to provide the first constraint of the number density of 21-cm absorbers at intermediate cosmological redshifts.

Throughout this paper we adopt the cosmological parameters $H_{0}=70$ km s$^{-1}$ Mpc$^{-1}$, $\Omega_{\Lambda}=0.7$ and $\Omega_{m}=0.3$. 

\subsection{ASKAP and BETA} 
ASKAP \citep{deboer09} is a wide-field aperture synthesis radio telescope that uses novel phased array feed technology (PAF) to provide a 30\,deg$^2$ field of view. The telescope is located at the radio-quiet Murchison Radio Observatory in Western Australia, and works in the frequency range 700-1800\,MHz. The full array of $36\times12$\,m ASKAP dishes came into operation in early 2019.  

The Boolardy Engineering Test Array \citep[BETA;][]{hotan14,mcconnell16} connected six ASKAP dishes equipped with first-generation (MkI) PAFs and operated from March 2014 to February 2016. Although BETA was originally intended purely as an engineering and commissioning testbed, some early science observations were also carried out \citep[e.g.][]{allison15,serra15,heywood16}.  The longest baseline for the BETA array was 916\,m, giving an angular resolution of about 1.8\,arcmin at 850 MHz. The frequency resolution was 18.5\,kHz, corresponding to a rest-frame velocity resolution of $\sim6$\,km\,s$^{-1}$.

In early observations with BETA, \cite{allison15}\ detected 21\,cm HI absorption at $z=0.44$ against a bright radio continuum source, PKS\,1740-517, whose spectroscopic redshift was unknown at the time. They found that the spectral dynamic range achievable with BETA was excellent for carrying out a blind absorption survey, and that the lowest ASKAP frequency band (covering 700-1000\,MHz), where the MkI PAFs were most sensitive, was essentially free of terrestrial radio-frequency interference (RFI). In spite of its limited collecting area, BETA therefore proved to be an excellent instrument with which to search for new HI absorption lines in the redshift range $0.4<z<1.0$ using bright ($>1$\,Jy) continuum sources as probes \citep[see e.g.][]{moss17,glowacki19}. 

The data presented in this paper were observed either with BETA, or during commissioning time with a 12-antenna ASKAP Early Science array (ASKAP-12) in early 2017. The ASKAP-12 array was equipped with the second-generation MkII PAF receivers (\citealt{chippendale15}). ASKAP-12 provided higher sensitivity than BETA, but with a more limited frequency coverage at this early stage (799--1039\,MHz, corresponding to $0.365<z<0.775$ for HI -- the 18.5\,kHz frequency resolution remained unchanged). 
The longest baseline for the ASKAP-12 array was around 2.3\,km, and the restoring beam for our data was $\approx 50 \times 25$\,arcsec (using Briggs' robust weighting = 0.5). 

\subsection{What do we expect to see?} 
Two kinds of 21\,cm absorption-line detections are possible: {\it associated}\ lines, where the neutral gas producing the absorption is within (or associated with) the radio source itself, and {\it intervening}\ lines, where the gas lies in a different galaxy somewhere along the line of sight to the radio source.  Associated absorption lines \citep[like the one seen in PKS\,1740-517 by][]{allison15} provide insights into the role of cold gas in AGN fuelling and feedback (see \citealt{morganti18} for a recent review). 

The focus of this paper is on intervening 21\,cm absorption lines, which have the potential to provide an independent probe of the cosmic HI mass density and its evolution. \cite{rao17} estimate the probability of intercepting a Damped Lyman-$\alpha$\ Absorber (DLA) with HI column density $N_{\rm HI} \geq 2 \times 10^{20} {\rm cm}^{-2}$ as: 
\begin{equation}
 n(z) = 0.027\pm0.007\ (1+z)^{1.682\pm0.200}.
\end{equation}
From this, we would expect $\sim$5\% of sightlines with $0.4<z<1.0$ (the HI redshift range spanned by the lowest-frequency ASKAP band at 700-1000\,MHz) to intersect an intervening DLA with an HI column density above this limit. 

How strong will these lines be? The observed HI 21\,cm optical depth ($\tau$) is related to the HI column density ($N_{\rm HI}$) by: 
\begin{equation}\label{equation:column_density}
N_{\rm HI} = 1.823\times10^{18}\,T_s \int \tau \Delta {V}, 
 \end{equation}
where $T_{\rm s}$ (in K) is the harmonic mean spin temperature of the HI gas and $\Delta$V the total absorption-line width in km\,s$^{-1}$. The optical depth of the absorber can be estimated from the observed spectrum, but also depends on the areal fraction of the radio source covered by the absorber (the covering factor $f$). If unknown, a fiducial value of $f = 1$ if often used, which is equivalent to estimating the optical depth, and hence $N_{\rm HI}$, as an average across the unresolved source. 

Although 21\,cm absorption-line measurements are particularly sensitive to colder gas, they do not provide a direct measurement of $N_{\rm HI}$ unless $T_{\rm s}$ is known or can be estimated. The spin temperature in \autoref{equation:column_density} is a column-density-weighted harmonic mean over all line-of-sight components of HI in the absorber. In the Milky Way ISM, the harmonic mean spin temperature is $T_{\rm s} \approx 300$\,K, with three HI phases in pressure equilibrium, consisting of cold (CNM; $T_{\rm s} \approx 100$\,K), unstable (UNM; $T_{\rm s} \approx 500$\,K) and warm neutral medium (WNM; $T_{\rm s} \approx 10^{4}$\,K), in mass fractions of 28, 20, and 52\,per\,cent, respectively (\citealt{murray18}). In general we expect that the mass fractions of these phases will vary depending on the physical conditions within each absorber. Cooling of the HI is driven by fine structure emission lines of C\,{\sc ii} and O\,{\sc i }, while heating occurs via the absorption of UV radiation by dust grains and subsequent photoelectric effect. The inferred spin temperature in a given 21-cm absorber is therefore dependent on the gas-phase metallicity, dust abundance and background UV field, and so likely to trace the evolutionary history of the galaxy. 

There are very few measurements of HI spin temperature available for neutral gas in the redshift range $0.4<z<1$.  \cite{ellison12} measured a spin temperature of $90\pm23$\,K at z=0.6 for an intervening line against the z=1.25 radio QSO\,J1431+3952, and \cite{zwaan15} derived a slightly lower value of $64\pm17$\,K for the same system.
\cite{kanekar14a} measured a range of spin temperatures from 90\,K to $>1380$\,K (median value $\sim270$\,K) for eight DLA systems in the redshift range $0.4<z<1$. Where specific measurements are unavailable, many authors assume a fiducial spin temperature of 100\,K, which equivalently corresponds to assuming that all of the absorbing gas is CNM. Using this value therefore gives a lower limit to the true column density of HI. 

In the absence of detailed spin temperature measurements, \cite{braun12} found a tight but strongly nonlinear relation between 21\,cm absorption opacity and HI column density in Local Group galaxies at $z=0$, and suggested that this relation might also be applicable at higher redshift.
For a DLA system with $N_{\rm HI} \sim 2\times10^{20}\,{\rm cm}^{-2}$, the \cite{braun12} relation gives a typical observed peak opacity of $\tau \sim0.015$. 

To probe DLA-like HI column densities in gas clouds with HI spin temperatures typical of galaxy disks, we ideally need to be able to detect 21\,cm absorption lines with a peak optical depth of $\tau \sim0.01-0.02$. To make reliable detections, we also require excellent spectral dynamic range across the telescope's bandwidth. 

\subsection{The expected locations of DLA systems at $0.4<z<1$}

Information on the host galaxies of $z<1.7$ DLA systems is relatively sparse \citep{wolfe05}. \cite{chen03} found that the galaxies that give rise to DLA systems at $z\leq1$\ span a wide range of morphological types, and that some DLAs may be associated with galaxy groups rather than individual galaxies.

On the theory side, simulations have presented contradictory results about where most of the HI is located, i.e. the relative contributions from the interstellar medium (ISM) of galaxies, the circum-galactic medium (CGM) and intergalactic gas as a function of cosmic time.  
\citet{van-de-voort12} showed that in the OWLS hydrodynamical simulations, the CGM starts to dominate the HI density at $z\gtrsim 1.5-2$, but other cosmological hydrodynamical simulations, find that the ISM HI continues to dominate the cosmic abundance out to higher redshift \citep{dave13}. In Illustris-TNG, \cite{diemer19} showed that the CGM becomes an important contributor of neutral hydrogen at $z\gtrsim 1$. 
On the other hand, a comparison of cosmological semi-analytic models of galaxy formation (which only account for HI in the ISM of galaxies) with inferred measurements of $\Omega_{\rm HI}$ has been used to argue that the CGM may already be important at $z\approx 1$ \citep{lagos14,lagos18}. 

The fact that some observations at intermediate redshifts, $0.3 \lesssim z \lesssim 1$, show DLAs to be associated with groups rather than individual galaxies \citep{peroux19, chen19} may indicate that the CGM is a significant reservoir of HI at these redshifts, in contrast to the local ($z\sim0$) Universe where most of the HI is observed to be in the ISM of galaxies. 
Given these current uncertainties, new observational constraints are likely to play a vital role in distinguishing between the various models. 

\ctable[  
notespar,
  star,
  caption={Targets observed with BETA},
  label={tab:sample_data},
  ]{l l lr rrrl l ll cc} 
{\tnote[]{Redshift references ({z$_{\rm ref}$}): 6dF=6dF Galaxy Survey \citep{jones09}; Al15=\cite{allison15}; dS94=\cite{di-serego94};  Ha03=\cite{halpern03}; He08=\cite{healey08}; Hu78=\cite{hunstead78}; Hu80=\cite{hunstead80}; Li99=\cite{lidman99}; Ma11=\cite{marshall11}; Mo78=\cite{morton78}; Mo82=\cite{morton82}; Mu84=\cite{murdoch84}; 
NTT=ESO NTT spectrum, this paper (see Appendix); 
Pe76=\cite{peterson76};  Pe79=\cite{peterson79}; Sh12=\citet{shaw12}; Ta93=\cite{tadhunter93}; Ti13=\cite{titov13}; Wh88=\cite{white88}; Wi83=\cite{wilkes83}; Wi00=\cite{wisotzki00}; Wr77=\cite{wright77}. \\ 
Position measurements are from the ICRF VLBI catalogue \citep{ma98} except for the following: PKS\,0903-57 position from \cite{murphy10}; MRC\,1039-474 from \cite{titov13} PKS\,1421-490 and MRC\,1613-586 from \cite{petrov11}; MRC\,1759-396 from \cite{fomalont03}; PKS\,1830-211 from \cite{fomalont00}, PKS\,2203-18 from \cite{beasley02}. \\ 
$^a$ For `in-band' radio sources with redshift $z\leq1.0$, we exclude a region within 3000 km\,s$^{-1}$ of the  emission redshift (column 8) since this region may be occupied by associated systems. \\
$^b$ See notes on individual objects in Appendix A.}}
{\FL 
\multicolumn{1}{c}{Name} & \multicolumn{1}{c}{RA} & \multicolumn{1}{c}{Dec} & \multicolumn{4}{c}{----- Radio flux density (Jy) ----- } & \multicolumn{1}{c}{$z_{\rm em}$}  &  \multicolumn{1}{c}{$z_{\rm ref}$} & \multicolumn{1}{c}{$\Delta z$}  & \multicolumn{1}{c}{Notes} \\
& \multicolumn{2}{c}{(J2000)} &\multicolumn{1}{c}{\small MRC} & \multicolumn{1}{c}{\small SUMSS} &\multicolumn{1}{c}{\small NVSS} &\multicolumn{1}{c}{\small AT20G} & & & probed$^a$ & \\
\multicolumn{1}{c}{(1)} & \multicolumn{1}{c}{(2)} & \multicolumn{1}{c}{(3)} & \multicolumn{1}{c}{(4)}  & \multicolumn{1}{c}{(5)} & \multicolumn{1}{c}{(6)} & \multicolumn{1}{c}{(7)} & \multicolumn{1}{c}{(8)}& \multicolumn{1}{c}{(9)}  & \multicolumn{1}{c}{(10)}  &  \multicolumn{1}{c}{(11)} \\ 
	\hline\hline
PKS\,0047-579 & 00:49:59.473 &  $-57$:38:27.34  &  2.52 & 2.05 & ...     & 1.87 &  1.797  & Pe76 & 0.60 & Background \\  
PKS\,0208-512 & 02:10:46.200 &  $-51$:01:01.89  &  5.48 & 3.37 & ...     & 3.29 &  1.003  & Pe76 & 0.60 & Background \\ 
PKS\,0302-623 & 03:03:50.631 &  $-62$:11:25.55  &  0.86 & 2.46 & ...     & 1.31 &  1.351  & He08  & 0.60 & Background  \\ 
PKS\,0438-43   & 04:40:17.180 &  $-43$:33:08.60  &  8.12 & 5.86 & ...     & 1.95  &  2.863  & Mo78 & 0.60 & Background  \\  
PKS\,0451-28   & 04:53:14.647 &  $-28$:07:37.33  &  2.14 &  ...    & 2.54 & 1.79  &  2.559   & Wi83 & 0.60 & Background \\  
\\
PKS\,0454-46   & 04:55:50.772 &  $-46$:15:58.68  &  4.25 & 2.87 & ...     & 4.16 &  0.853   & Wi00  & 0.43 & In-band \\    
PKS\,0506-61   & 05:06:43.989 &  $-61$:09:40.99  &  5.03 & 3.12 & ...     & 1.66 & 1.093    & Wr77  & 0.60 & Background \\  
PKS\,0537-441 & 05:38:50.362 &  $-44$:05:08.94  &  2.56 & 3.52 & ...     & 5.29 & 0.894 & Pe76 & 0.47 & In-band \\  
PKS\,0637-75   & 06:35:46.508 &  $-75$:16:16.82  &  7.89 & 5.38 & ...     & 3.14 & 0.653 & Hu78 & 0.23 & In-band \\  
PKS\,0743-67   & 07:43:31.612 &  $-67$:26:25.55  &  8.61 & 5.68 & ...     & 1.22 & 1.512  & dS94 & 0.60 & Background \\ 
 \\
 PKS\,0903-57  & 09:04:53.36   &  $-57$:35:04.7    &  4.90 & 3.31 & ...     & 1.43 & ...    & ..  & ($\leq0.6$) & Uncertain$^b$ \\ 
 PKS\,0920-39  & 09:22:46.418 &  $-39$:59:35.07  &  4.38 & 3.16 & 2.62 & 1.31 & 0.591    & Wh88 & 0.17 & In-band  \\ 
 MRC\,1039-474 & 10:41:44.650 &  $-47$:40:00.06  &  1.44 & 2.37 & ...    & 1.26 & 2.558  & Ti13  & 0.60 & Background$^b$  \\ 
 PKS\,1104-445 & 11:07:08.694 &   $-44$:49:07.62 &  1.49 & 2.52 & ...    & 1.67 & 1.598 & Pe79 & 0.60 & Background   \\   
 PKS\,1421-490 & 14:24:32.237 &  $-49$:13:49.74  & 13.10& 9.68 & ...    & 2.64 & 0.662  & Ma11 & 0.24 & In-band  \\ 
 \\
 PKS\,1424-41   & 14:27:56.298 &  $-42$:06:19.44  &  6.39 & 3.86 & ...    &  2.74 & 1.522 & Wh88 & 0.60 & Background  \\  
 PKS\,1504-167 & 15:07:04.787 &  $-16$:52:30.27  &  2.20 & ...    & 2.71 & 1.05  & 0.876 & Hu78 & 0.45 & In-band   \\	
 MRC\,1613-586 & 16:17:17.889 & $-58$:48:07.86  &  3.12 & 3.10 & ...    & 2.71  & 1.422   & Sh12   & 0.60 & Background \\  
 PKS\,1610-77   &  16:17:49.276 & $-77$:17:18.47  &  5.35 & 4.15 &  ...   & 1.86 & 1.710  & Hu80 & 0.60 & Background  \\ 
 PKS\,1622-253 &  16:25:46.892 & $-25$:27:38.33  &  2.36 & ...    & 2.52 & 2.06 & 0.786  & dS94 & 0.35 & In-band  \\ 
 \\
PKS\,1622-29    &  16:26:06.021 & $-29$:51:26.97 &  2.74  & ...    & 2.29 & 1.79 & 0.814 & NTT & 0.39  & In-band$^b$ \\ 
PKS\,1740-517 & 17:44:25.451  & $-51$:44:43.79  &  5.38 &  8.15 & ...    &  1.24  & 0.441 & Al15  & 0.02 & In-band$^b$  \\ 
MRC\,1759-396 & 18:02:42.680 &  $-39$:40:07.90 &  2.56 &  1.38 & 2.27 & 1.41 & 1.319  & Sh12  & 0.60 & Background$^b$ \\  
PKS\,1830-211  & 18:33:39.886  & $-21$:03:40.57 & 11.47 & ...     & 10.90 &  5.50 & 2.507 & Li99 & 0.60 & Background \\  
MRC\,1908-201  & 19:11:09.653  & $-20$:06:55.11  &  1.94 & ...     &  2.71  &  2.67 & 1.119 & Ha03 & 0.60 & Background  \\ 
\\
MRC\,1920-211 & 19:23:32.190  & $-21$:04:33.33  & 1.35  & ...     &  3.17 &  2.55 &  0.874 & Ha03 & 0.45 & In-band \\ 
PKS\,2052-47    & 20:56:16.360  & $-47$:14:47.63 & 4.15  & 2.14 &  ...     &  1.17 &  1.492  & Mu84 & 0.60 & Background \\ 
PKS\,2155-152  & 21:58:06.282   & $-15$:01:09.33 & 2.51 &  ...     & 3.02 &  1.90 &  0.672 & Wh88 & 0.25 & In-band  \\ 
PKS\,2203-18    & 22:06:10.417   & $-18$:35:38.75 & 9.73 & ...      & 6.40 &  2.03 &  0.619 & Mo82 & 0.19 & In-band \\ 
PKS\,2326-477  & 23:29:17.704   & $-47$:30:19.12 & 3.21 & 3.05  & ...     &  1.42 &  1.304 & 6dF   & 0.60 & Background  \\ 
\\
PKS\,2333-528  & 23:36:12.145   & $-$52:36:21.95 &  1.81 & 2.16 & ...     &   1.07 & ...   & ..   & ($\leq0.6$) & Uncertain  \\  
PKS\,2345-16     & 23:48:02.609   & $-$16:31:12.02 &  2.09 & ...    & 2.64  &  2.45 &  0.576 & Ta93 & 0.15 & In-band   \\	
\\
\multicolumn{10}{l}{Total path length for the 30 sources of known redshift: $\Delta z ({\small {\rm BETA}}) = 14.00$} \\
\LL}

\ctable[
  notespar, 
  star,
  cap = {Main data table (2)},
  caption={Targets observed with ASKAP-12},
  label={tab:sample_data2},
  ]{l l lr rrrl l ll cc}%
{\tnote[]{Redshift references ({$z_{\rm ref}$}): 6dF=6dF Galaxy Survey \citep{jones09}; Ar67=\cite{arp67}; Dr97=\cite{drinkwater97} Fr83=\cite{fricke83};  Ha03=\cite{halpern03}; 
He10=\cite{hewett10}; 
Hu80=\cite{hunstead80}; Li99=\cite{lidman99}; 
Ma96=\cite{marziani96}; 
O'D91=\cite{odea91}; Os94=\cite{osmer94}; 
St74=\cite{strittmatter74};
St89=\cite{stickel89}; Wh88=\cite{white88}; Wi83=\cite{wilkes83}; Wi86=\cite{wilkes86}; 
Wr83=\cite{wright83}. \\
Position measurements are from the ICRF VLBI catalogue \citep{ma98} except for the following: 
PKS\,0122-00 and PKS\,1136-13 from \cite{beasley02}; PKS\,1229-02, PKS\,1245-19, PKS\,1508-05, PKS\,2123-463 and PKS\,2244-37 from \cite{fey15}.  
\\ 
$^a$ For `in-band' radio sources with redshift $z\leq0.77$, we exclude a region within 3000 km\,s$^{-1}$ of the emission redshift (column 8) since this region may be occupied by associated systems. \\
$^b$ See notes on individual objects in Appendix A. \\
$^c$ For PKS\,1229-02 and PKS\,1508-05, the redshift path probed was reduced slightly due to correlator errors over part of the observed frequency range.  }
}
{\FL
 \multicolumn{1}{c}{Name} & \multicolumn{1}{c}{RA} & \multicolumn{1}{c}{Dec} & \multicolumn{4}{c}{----- Radio flux density (Jy) ----- } & \multicolumn{1}{c}{$z_{\rm em}$ }& \multicolumn{1}{c}{$z_{\rm ref}$} & \multicolumn{1}{c}{$\Delta z$} & \multicolumn{1}{c}{Notes} \\
& \multicolumn{2}{c}{(J2000)} &\multicolumn{1}{c}{\small MRC} & \multicolumn{1}{c}{\small SUMSS} &\multicolumn{1}{c}{\small NVSS} &\multicolumn{1}{c}{\small AT20G} & & & probed$^a$ \\
\multicolumn{1}{c}{(1)} & \multicolumn{1}{c}{(2)} & \multicolumn{1}{c}{(3)} & \multicolumn{1}{c}{(4)}  & \multicolumn{1}{c}{(5)} & \multicolumn{1}{c}{(6)} & \multicolumn{1}{c}{(7)} & \multicolumn{1}{c}{(8)}& \multicolumn{1}{c}{(9)}  & \multicolumn{1}{c}{(10)}  &  \multicolumn{1}{c}{(11)} \\ 
\hline \hline
\multicolumn{5}{l}{(a) New targets} \\
PKS\,0122-00   & 01:25:28.844  & $-00$:05:55.93   &  1.20 & ...    &  1.54 & 1.16 & 1.075   &  6dF & 0.40 & Background \\   
PKS\,0237-23   & 02:40:08.175  & $-23$:09:15.73   &  3.67 & ...    &  6.26 & 0.90 &  2.223  & Ar67 & 0.40 & Background \\  
PKS\,0405-12   & 04:07:48.431  & $-12$:11:36.66   &  8.17 & ...    & 2.94  & 1.25  & 0.573 &  Ma96 & 0.18 & In-band        \\ 
PKS\,0454-234 & 04:57:03.179  & $-23$:24:52.02   &  ...    & ...    & 1.73  & 3.84  & 1.003 &  St89 & 0.40 & Background \\   
PKS\,0458-02   & 05:01:12.810  & $-01$:59:14.26   & 2.30  & ...   &  2.26  & 1.10  & 2.286 &  St74 & 0.40  & Background \\
 &&&&& \\
PKS\,0805-07   & 08:08:15.536  & $-07$:51:09.89   & 2.49  & ...   &  1.60  & 0.77  & 1.837 & Wh88  & 0.40 & Background \\  
PKS\,0834-20   & 08:36:39.215  & $-20$:16:59.50   & 3.53  & ...   &  1.97  & 2.67  & 2.752 & Fr83  & 0.40 & Background  \\ 
PKS\,0859-14   & 09:02:16.831  & $-14$:15:30.88   & 3.93  & ...   &  2.90  & 1.04  & 1.332 &  6dF  & 0.40 & Background$^b$  \\ 
PKS\,1127-14   & 11:30:07.053  & $-14$:49:27.39    &  5.07 & ...    & 5.62  & 1.87  & 1.184 &  Wi83 & 0.40 & Background \\ 
PKS\,1136-13   & 11:39:10.703  & $-13$:50:43.64    & 10.50 & ...   & 4.22  & 0.54  & 0.556 &  6dF   & 0.16 & In-band       \\ 
&&&&& \\
PKS\,1144-379  & 11:47:01.371  & $-38$:12:11.02     &   0.94 & 0.81 & 1.80 & 1.38 & 1.048 &  St89  & 0.40 & Background \\   
PKS\,1229-02   &  12:32:00.016  & $-02$:24:04.80     &   3.87 & ...   & 1.65  &  0.90  & 1.045 &  He10  & 0.27$^c$ & Background \\ 
PKS\,1245-19   &  12:48:23.898  & $-19$:59:18.59     &   8.61 & ...   &  5.14 &  0.69  & 1.275 & O'D91  & 0.40 & Background$^b$ \\ 
PKS\,1508-05   &  15:10:53.592  & $-05$:43:07.42     &   7.71 & ...   &  3.57 &  1.27 & 1.185 &  Wi86   & 0.36$^c$ &  Background \\
PKS\,1935-692  & 19:40:25.528 & $-69$:07:56.97     &  1.70  &  1.75 & ...   &  0.52  &  3.154 & Os94  & 0.40 & Background \\ 
&&&&& \\
PKS\,2106-413  & 21:09:33.189 & $-41$:10:20.61     &  2.66  &  1.82 & ...   &  1.63  & 1.058  & Wh88  & 0.40 & Background   \\  
PKS\,2123-463  & 21:26:30.704 & $-46$:05:47.89     &  1.95  &  1.52 & ...   &  0.55  & ...   &  ..    & ($\leq0.4$) & Uncertain$^b$ \\  
PKS\,2131-021  & 21:34:10.310 & $-01$:53:17.24     & 1.91   &   ...    & 1.69 & 2.11 & 1.285  & Dr97   & 0.40 & Background   \\  
PKS\,2204-54    & 22:07:43.733 & $-53$:46:33.82     & 2.55   &  1.80  &  ...  &  1.12 & 1.215  & Wi83 & 0.40 & Background   \\  
PKS\,2223-05   & 22:25:47.259  & $-04$:57:01.39    & 11.89 & ...       & 7.41 &  8.32 &  1.404  &  Wr83  & 0.40 & Background$^b$ \\  
&&&&&  \\
PKS\,2244-37   & 22:47:03.917  & $-36$:57:46.30    &  2.95 &  1.71 & 1.26 &   0.94 &   2.252 &  Wi83  & 0.40 & Background \\   
&&&&& \\
\multicolumn{10}{l}{Total path length for the 20 sources of known redshift: $\Delta z$({\small {\rm ASKAP-12}}) = 7.37} \\
&&&&& \\ 
\multicolumn{5}{l}{(b) Repeat observations of BETA targets} \\
PKS\,1610-77   &  16:17:49.276 & $-77$:17:18.47  &  5.35 & 4.15 &  ...   & 1.86 & 1.710   & Hu80  & 0.40 & Background  \\ 
PKS\,1830-211  & 18:33:39.886  & $-21$:03:40.57 & 11.47 & ...     & 10.90 &  5.50 & 2.507 & Li99 & 0.40 & Background \\ 
MRC\,1908-201 & 19:11:09.653  & $-20$:06:55.11  &  1.94 & ...    & 2.71 &  2.67  & 1.119 &  Ha03 & 0.40 & Background  \\ 
 \LL
}

\section{Sample selection}
For ease of interpretation in this pilot study, we aimed to select background radio sources that were both bright and reasonably compact (i.e. had one or more radio components that were unresolved on arcsec scales). 

We used the Australia Telescope 20\,GHz (AT20G) Bright Source Sample (BSS) catalogue \citep{massardi08} as the basis for our initial BETA target selection, since it contains a high fraction of radio-loud QSOs (with a median redshift of $z\sim1.2$) and so is dominated by distant, compact radio sources. This catalogue covers the whole sky south of declination $-15^\circ$ (apart from a small strip with Galactic latitude $|b|<1.5^\circ$). For our later observations with ASKAP-12, we supplemented the BSS catalogue with more northerly objects from the main AT20G source catalogue \citep{murphy10}. 

\subsection{Target selection for BETA} 
We selected the initial pilot sample of bright radio continuum sources to observe with BETA based on two main considerations: 
\begin{enumerate}
\item
The limited sensitivity of the BETA telescope means that the background radio sources used as probes need to be bright enough (ideally with flux density $>2$\,Jy at 700-1000\,MHz) to allow us to measure lines with optical depth $\tau\sim0.01$, and to distinguish weak absorption lines from noise fluctuations in a reliable way. 
\item 
The background sources should ideally be at redshift $z>1$ to allow us to probe the full $\Delta z=0.6$ redshift path length available with BETA, but we also included bright radio sources for which no optical redshift is currently available since these may be objects where a dusty galaxy intervenes along the line of sight to a distant radio source. 
\end{enumerate} 

We began by selecting the 130 AT20G BSS objects with 20\,GHz flux density above 1.0\,Jy (38 of these have 20 GHz flux densities above 2.0\,Jy), and removed Galactic sources and other objects known to have redshift $z<0.4$. This left 112 AT20G BSS sources.  
We then further restricted the list to objects with an NVSS or SUMSS flux density above 2.0\,Jy (to ensure good S/N with BETA). 

This left us with the final sample of 32 sources listed in Table \ref{tab:sample_data}.  Two of these objects lack a reliable optical redshift. Of the remaining 30 sources, 17 are background sources for the whole ASKAP band (i.e. have redshift $z>1$) and 13 have redshifts that place the HI line within the lowest ASKAP band ($0.4<z<1.0$).
For the 13 `in-band' radio sources with redshift $z\leq1.0$, we reduced the assumed redshift path by $\Delta z=0.01$ to exclude the region within 3000 km\,s$^{-1}$ of the emission redshift that may be occupied by associated HI absorption systems. 
After accounting for this `proximity effect', the total redshift interval probed for intervening DLA systems by the 30 sources of known redshift is $\Delta z=14.00$.

\subsection{Target selection for ASKAP-12} 
We selected some additional bright, compact sources to observe with ASKAP-12 during commissioning time in February 2017. The improved sensitivity of this 12-antenna array allowed us to relax some of the constraints on our earlier BETA sample, and the new sources were chosen from the AT20G catalogue \citep{murphy10} as follows: 

\begin{enumerate}
\item
We relaxed the $-15^\circ$ declination limit applied for the BETA sample, and included sources up to dec $0^\circ$. 
\item
As before, we excluded Galactic sources and other objects known to have redshift $z<0.4.$
\item 
We include all the remaining objects with 20\,GHz flux density above 0.5\,Jy (rather than 1.0\,Jy for the BETA sample) and NVSS or SUMSS flux density above 1.5\,Jy (rather than 2.0\,Jy for the BETA sample). 
\end{enumerate} 

This left us with the sample of 21 additional sources listed in Table \ref{tab:sample_data2}.  Twenty of these objects have a reliable optical redshift, 18 are background sources for the whole ASKAP band (i.e. have redshift $z>1$) and two have redshifts that place the HI line within the lowest ASKAP band ($0.4<z<1.0$). 
As for our BETA sample, we reduced the assumed redshift path for the two `in-band' sources by $\Delta z=0.01$ to account for proximity effects. 
The total redshift interval probed for intervening DLA systems by the sources in Table \ref{tab:sample_data2} is $\Delta z=7.37$. 
Two sources observed with BETA were re-observed with ASKAP-12, and these are also listed in Table \ref{tab:sample_data2}.  

\subsection{Structure of the target radio sources}
Our target sources were selected to be compact, and in most cases we expect their radio emission to be dominated by a single component with angular size smaller than 1\,arcsec. We can assess this in several ways: 

\begin{itemize}
\item 
{\bf ATCA calibrators:}\ 50 of the 53 sources in Tables 1 and 2 are ATCA calibrator sources\footnote{The three objects not listed as ATCA calibrators are PKS\,0743-67, PKS\,1229-02 and PKS\,2123-463} listed in the online calibrator database.\footnote{\url{http://www.narrabri.atnf.csiro.au/calibrators/calibrator_database.html}} 
For these sources, we can use the `defect' and `closure phase' parameters listed in the calibrator database to identify any sources that have resolved emission on arcsec scales at frequencies above 1.4\,GHz.  
\item
{\bf AT20G 6-km visibilities:}\ 48 of the 53 sources have visibility measurements at 20\,GHz on the longest (6\,km) ATCA baselines \citep{chhetri13}, which allow us to assess the compactness of sources on scales of $\sim150$\,mas. 
\item
{\bf VLBI images:} 36 of the sources in Tables 1 and 2 have VLBI images from \cite{ojha05,ojha10} or \cite{pushkarev17} that map out the structure of compact components on scales as small as 1-10\,mas. 
\end{itemize}

While these data are useful, they provide at best an imperfect picture of the structure of these sources in the frequency range observed by ASKAP. As noted by \cite{kanekar14a}, we would ideally like to have lower-frequency VLBI images that allow us to estimate what fraction of the 700-1000\,MHz radio emission seen by ASKAP (which had $\sim1$\,arcmin resolution at the time of our commissioning observations) is located in compact components less than 10--20\,mas in angular size. 

Table \ref{tab:sample_res} lists galaxies in our sample that are known to have some resolved continuum emission on arcsec scales (though each of these objects also has a strong compact core). These objects were identified either because the ATCA calibrator database showed evidence for structure on arcsecond scales, or because the \cite{chhetri13} 6-km visibility had a value $<0.9$, indicating the presence of resolved emission on scales larger than about 0.15\,arcsec.  We then carried out a literature search for arcsec-scale radio images of these objects, and the results are summarized in column (5) of Table \ref{tab:sample_res}. 
PKS\,1830-211, also listed in this table, is a special case since the resolved emission is the result of gravitational lensing by a foreground galaxy, rather than being intrinsic to the source. 

Most of the sources with published VLBI images at 5--15\,GHz  \citep{ojha05,ojha10,pushkarev17,petrov19}
appear to have a significant fraction (typically 20-60\%) of their flux density in components smaller than $\sim10$\,mas. This corresponds to an angular scale of roughly 50\,pc at $z\sim0.4$, 70\,pc at $z\sim0.7$ and 80\,pc at $z\sim1.0$, which is slightly smaller than the expected size of individual HI clouds in galaxy disks \citep[estimated as $\sim100$\,pc; ][]{braun12}. Even without a more detailed knowledge of the source structure, therefore, it appears that most of our target sources remain quite compact on scales as small as 10-20\,mas, making them effective probes for intervening HI absorption-line systems (because there is less need to account for a non-unity covering factor $f$). 

This conclusion is supported by the work of \cite{horiuchi04}, who observed a large sample of powerful flat-spectrum radio AGN at 5\,GHz with both the VLBA and VSOP. They found that a typical AGN in their sample had about 50\% of its radio emission in a component smaller than 10\,mas in size, with around 40\% of this milliarcsec-scale emission (20\% of the total emission) coming from a radio core with an average size of 0.2\,mas.

\ctable[  
notespar,
  star,
  caption={Sources known to have resolved continuum emission on arcsec scales. Column 4 lists the 20\,GHz visibilities on the longest (6-km) ATCA baseline, from \protect\cite{chhetri13}. Sources with 6k\_vis$\geq0.9$ are expected to have almost all their high-frequency radio emission originating from a region less than about 120\,mas in angular size. },
label = {tab:sample_res},
  ]{l l ll llll l ll cc} 
{\tnote[]{References: Hi83=\cite{hintzen83}; Ja91=\cite{jauncey91}; Kr92=\cite{kronberg92};  Ma05=\cite{marshall05}; Mc12=\cite{mcconnell12};  Sa89=\cite{saikia89}; Sa04=\cite{sambruna04} }}
{\FL 
\multicolumn{1}{l}{Name} & \multicolumn{1}{c}{$z_{\rm ref}$} 
& \multicolumn{1}{c}{Cal?} 
& \multicolumn{1}{c}{6k\_vis} & \multicolumn{1}{c}{Notes} & \multicolumn{1}{c}{Ref.} \\
\multicolumn{1}{c}{(1)} & \multicolumn{1}{c}{(2)} & \multicolumn{1}{c}{(3)} & \multicolumn{1}{c}{(4)}  & \multicolumn{1}{c}{(5)} & \multicolumn{1}{c}{(6)} \\
	\hline\hline
PKS\,0405-12 & 0.573 & C & 0.94 & Triple at 1.4\,GHz, strong core, LAS$\sim40$\,arcsec & Sa04 \\
PKS\,0903-57 & .. & C & 0.69 & Compact double at 5\,GHz, LAS $\sim3$\,arcsec & Ma05 \\
PKS\,1136-13 & 0.556 & C & 0.45 & Triple at 1.4\,GHz, strong core, LAS$\sim20$\,arcsec & Sa04 \\
PKS\,1229-02 & 1.045 & .. & 0.89 & Triple at 1.6\,GHz, LAS $\sim20$\,arcsec & Kr92 \\
PKS\,1421-490 & 0.662 & C & 0.76 & Strong core at 5\,GHz with $\sim1$\,arcsec jet & Ma05 \\
PKS\,2123-463 & .. &  ... & 0.91 & Two components at 5\,GHz, LAS$\sim4$\,arcsec & Mc12 \\
&&& \\
PKS\,1830-211 & 2.507 & C & 0.19 & Lensed double/ring at 1.7\,GHz, LAS$\sim1$\,arcsec & Ja91 
\LL }

\section{Observations} 
The observational techniques and data reduction used were similar to those described by \cite{allison15} and \cite{allison17}, and we refer the reader to those papers for further details.

\subsection{Observations with BETA} 
We observed the 32 objects listed in Table \ref{tab:sample_data} with BETA over the period from July 2014 to February 2016. The total integration time was typically 3-5 hours for each object, with between 4 and 6 BETA dishes in the array.  The BETA telescope was used as an engineering testbed throughout its operation, so there were sometimes technical problems that made part or all of an observation unusable. When this occurred, the observation was repeated until good-quality data were obtained. 

In principle up to nine beams could be formed for wide-field imaging using the Mark I phased array feeds (PAFs) on BETA, positioned anywhere within the 30 deg$^2$ field of view of the PAF. However, for our targeted observations presented here, we used a single PAF beam centred at the position of the target source. To obtain initial solutions for the complex antenna gains and to calibrate the flux density scale \citep[based on the model of][]{reynolds94}, we accompanied each observation with a short integration on PKS\,B1934$-$638 between 5 and 15\,min. The expected uncertainty in the flux density scale is 2--3\% \citep{heywood16}. 

Our observations with BETA were carried out exclusively using the lower frequency band, between 711.5 and 1015.5\,MHz, equivalent to \mbox{H\,{\sc i}} redshifts between $z_{\rm HI} = 0.4$ and $1.0$. The fine channelization generated 16,416 channels across the 304\,MHz bandwidth, with an effective spectral resolution between approximately 5.5 and 7.8\,km\,s$^{-1}$. The full width at half power of the PAF beams is approximately 1.7 degrees at the band centre, and the spatial resolution of BETA is approximately 1\,arcminute (using uniform weighting) so that we do not expect any of our objects to be spatially resolved.

\subsection{Observations with ASKAP-12}  
We observed the 23 objects listed in Table \ref{tab:sample_data2} with ASKAP-12 during January--February 2017. The total integration time was typically 2 hours for each object. ASKAP-12 usually had between 12 and 14 antennas operational for each observation, each fitted with Mark II PAFs with improved sensitivity at 1400\,MHz \citep[see][]{chippendale15}. 

Up to 36 PAF beams could be electronically formed to fully sample the 30 deg$^2$ field of view. However, for our observations we only considered a single beam centred on the object source. Since calibration of the complex antenna gains and flux scale in each PAF beam would require a separate observation of PKS\,1934-638, only forming a single beam greatly improves observing efficiency for this work. Secondly, during commissioning observations the ASKAP-12 backend capacity was limited so that forming a single PAF beam allowed us to achieve a larger spectral bandwidth. 

We observed simultaneously at all frequencies between 799.5 and 1039.5\,MHz, spanning \mbox{H\,{\sc i}} redshifts between $z_{\rm HI} = 0.37$ and 0.77. The fine channelisation produced 12960 channels across the 240\,MHz, with an effective spectral resolution between 5.3 and 6.9\,km\,s$^{-1}$. The PAF beam full width at half power at the band centre is approximately 1.6\,degrees and the spatial resolution is approximately 30\,arcsec, again meaning that none of our targets are expected to be spatially resolved by these observations.
 
\subsection{Data reduction} 
 
The ingested data from the ASKAP correlator were recorded in measurement set format and so initial flagging (autocorrelations and amplitude thresholding) and splitting of the data were performed using the \textsc{CASA} package \citep{mcmullin07}. Subsequent automated flagging, calibration, imaging and continuum subtraction of the data were carried out using the \textsc{MIRIAD} package \citep{sault95}. 
 
 The full-spectral-resolution visibilities were split into sub-band chunks, so as to enable efficient parallelization of the data processing and to remove spectral discontinuities caused by the PAF beams. The PAF beams were formed electronically in fixed frequency intervals; every 4 or 5\,MHz for BETA and 1\,MHz for ASKAP-12. These generated discontinuous jumps in the complex gain response of the telescope as a function of frequency and therefore needed to be corrected during bandpass calibration and refined further in continuum subtraction. In the case of BETA, these intervals equate to velocities greater than 1000\,km\,s$^{-1}$ and are therefore much larger than the typical linewidths expected for absorption. We corrected for this simply by splitting the data into sub-band chunks and performing bandpass calibration and continuum subtraction individually on these data.

However, in the case of the 1\,MHz intervals used in ASKAP-12 this approach could lead to removal of absorption lines wider than 300\,km\,s$^{-1}$ during continuum subtraction. We therefore split the ASKAP-12 data into sub-band chunks of 4\,MHz (i.e. four beamforming intervals). To correct for the discontinuities in ASKAP-12 data that occur every 1\,MHz, we solved for the bandpass per channel using PKS\,1934-638 and then recovered S/N by using GPEDIT to smooth the solutions using a 10-channel Hanning window with a break every 1\,MHz. Outliers in the bandpass solutions, which are generated either by hardware glitches or radio frequency interference, were identified using the interquartile range and replaced through interpolation before smoothing.
 
 Separately, a single full-bandwidth data set, averaged to 1\,MHz resolution, was used to obtain high signal-to-noise continuum images for self-calibration. Initial solutions to the antenna gains as a function of time were obtained using a sky model based on the catalogues of SUMSS \citep{mauch03}, MGPS2 \citep{murphy07} and NVSS \citep{condon98}. Further iterative refinement of the solutions to smaller time intervals were then carried out using self-calibration based on the imaged continuum data. These gain solutions were then applied to the full-spectral-resolution data in each sub-band chunk.

Continuum subtraction was carried out separately on each sub-band chunk; first using the CLEAN algorithm to generate a continuum model, which was then subtracted from the visibilities using UVMODEL, followed by UVLIN to fit and subtract a second order polynomial from the residuals. Data cubes were formed by imaging the continuum-subtracted visibilities in each sub-band chunk and a single spectrum was constructed at the position of peak emission from the continuum source. The continuum flux density was also measured at the same position in each sub-band chunk, so that the fractional absorption could be accurately calculated as a function of frequency. In cases where an object was observed on multiple occasions we formed a single spectrum by carrying out an inverse-variance-weighted average.

\subsection{Data tables}
Tables \ref{tab:beta_obs} and \ref{tab:askap12_obs} list some key parameters of the objects observed with BETA and ASKAP-12 respectively, arranged as follows: \vspace{0.1cm} \\
\noindent
(1) Source name from Table \ref{tab:sample_data} or \ref{tab:sample_data2}\\
(2) Total observing time (in hours) \\
(3) Mean continuum flux density of the source (in mJy) averaged across the full continuum frequency band observed \\
(4) Standard deviation of the mean flux density - for these strong sources, this is a mainly a measure of how much the continuum flux density varies across the 240--300 MHz band \\ 
(5) Typical (median) rms noise (in mJy) in a single spectral channel \\
(6) Typical 1$\sigma$ sensitivity in optical depth for a single spectral channel. \\

\ctable[
notespar,
cap = {BETA obs},
caption = {Measurements from BETA observations of the target sources in Table \ref{tab:sample_data}.  See \S3.4 of the text for a description of each column. For objects with more than one observation, the values listed are for the spectrum with the best optical-depth sensitivity $\sigma_{\tau}$. },
label = {tab:beta_obs}
]{lcrrclr r}%
{ 
}{
\FL
\multicolumn{1}{c}{Name} &  \multicolumn{1}{c}{t}  &\multicolumn{1}{c}{$S_{\rm cont}$} & \multicolumn{1}{c}{$\Delta$S} & \multicolumn{1}{c}{rms/ch} & \multicolumn{1}{c}{$\sigma_{\tau}$}  \\
  &  \multicolumn{1}{c}{(h)} & \multicolumn{1}{c}{mJy} &   \multicolumn{1}{c}{mJy} & \multicolumn{1}{c}{mJy} &   \multicolumn{1}{c}{}  \\  
\multicolumn{1}{c}{(1)} & \multicolumn{1}{c}{(2)} & \multicolumn{1}{c}{(3)} & \multicolumn{1}{c}{(4)}  & \multicolumn{1}{c}{(5)} &  \multicolumn{1}{c}{(6)}  \\
\hline \hline
PKS\,0047-579 &  5  & 2198 &    81   & 21.7  & 0.0099 \\ 
PKS\,0208-512 &  3  & 2748 &  163   & 30.3  & 0.0110 \\ 
PKS\,0302-623 &  5  &  4426 &  429   & 43.0  & 0.0097 \\
PKS\,0438-43   &  3  &  5455 &  354   & 37.2  & 0.0068 \\
PKS\,0451-28   &  5  &  2430 &   94    & 23.4  & 0.0096 \\ 
\\
PKS\,0454-46   &  5  & 3680 & 130 & 22.9 & 0.0062 \\
PKS\,0506-61   & 3.5 & 3154 & 263 & 31.0 & 0.0098 \\
PKS\,0537-441 &  5  & 4318 & 183 & 22.8 & 0.0053 \\
PKS\,0637-75   &  4  & 6250 & 663 & 26.4 & 0.0042 \\ 
PKS\,0743-67   &  4  & 5276 & 407 & 24.2 & 0.0046 \\
\\
 PKS\,0903-57  &  3  & 3156 & 194 & 28.4 & 0.0090 \\
 PKS\,0920-39  &  5  & 3233 & 152 & 25.5 & 0.0079 \\
 MRC\,1039-474 & 5 & 1867 & 77 & 24.4 & 0.0130 \\
 PKS\,1104-445 &  4 & 2417 & 502 & 25.0 & 0.0103\\
 PKS\,1421-490 &  2 & 9360 & 512 & 36.3 & 0.0039 \\
 \\ 
 PKS\,1424-41   & 3.5 & 4714 & 153 & 27.9 & 0.0059 \\
 PKS\,1504-167 &  5 & 1634 &  73  & 23.6 & 0.0144 \\
 MRC\,1613-586 & 5 & 4586 &104 & 24.4 & 0.0053 \\
 PKS\,1610-77   &  5 & 3933 & 206 & 16.5 & 0.0042 \\ 
 PKS\,1622-253 &  5 & 2222 & 102 & 29.5 & 0.0133 \\ 
 \\
PKS\,1622-29    &  5  & 2755 & 115  & 23.7 & 0.0086 \\
PKS\,1740-517  &      &   \\
MRC\,1759-396 &  4  & 1489 &   29 &  29.0 & 0.0195 \\
PKS\,1830-211  &  3.5  & 12620 & 521 & 37.6 & 0.0030 \\
MRC\,1908-201  &  5  &  1768 &   57 & 23.7 &  0.0134 \\
\\
MRC\,1920-211 &  5  &  2331 & 175 & 24.8 & 0.0106 \\
PKS\,2052-47    &  5  & 2523  & 140 & 22.1 & 0.0088 \\
PKS\,2155-152  &  5  & 4096  &   88 & 24.9 & 0.0061 \\
PKS\,2203-18    &  3   & 7395  & 416 & 58.0 & 0.0078 \\
PKS\,2326-477  &  4  & 4169  & 151 & 35.3 & 0.0085 \\
\\
PKS\,2333-528  &  5  & 2233  &  58  & 26.8 & 0.0120 \\
PKS\,2345-16    &  5  & 2684  &  39  & 27.4 & 0.0102 
 \LL
}

\ctable[
notespar,
cap = {ASKAP-12 obs},
caption = {Measurements from ASKAP-12 observations of the target sources in Table \ref{tab:sample_data2} .  See \S3.4 of the text for a description of each column. },
label = {tab:askap12_obs}
]{lcrrclr r}%
{ 
}{
\FL
\multicolumn{1}{c}{Name} &  \multicolumn{1}{c}{t}  &\multicolumn{1}{c}{$S_{\rm cont}$} & \multicolumn{1}{c}{$\Delta$S} & \multicolumn{1}{c}{rms/ch} & \multicolumn{1}{c}{$\sigma_{\tau}$}  \\
  &  \multicolumn{1}{c}{(h)} & \multicolumn{1}{c}{mJy} &   \multicolumn{1}{c}{mJy} &   \multicolumn{1}{c}{mJy} &   \multicolumn{1}{c}{} \\  
\multicolumn{1}{c}{(1)} & \multicolumn{1}{c}{(2)} & \multicolumn{1}{c}{(3)} & \multicolumn{1}{c}{(4)}  & \multicolumn{1}{c}{(5)} &  \multicolumn{1}{c}{(6)}\\
\hline \hline
PKS\,0122-00   & 2 & 1633 & 75 & 12.6 & 0.0078\\    
PKS\,0237-23   & 2 & 6824 & 258 & 12.6 & 0.0018 \\  
PKS\,0405-12   & 2 & 2638 & 50 & 12.5 & 0.0044 \\
PKS\,0454-234  & 2 & 2020 & 10 & 12.3 & 0.0061 \\    
PKS\,0458-02   & 2 & 1376 & 13 & 15.1 & 0.0110 \\ 
\\
PKS\,0805-07   & 2 & 1895 & 27 & 13.6 & 0.0072 \\  
PKS\,0834-20   & 2 & 2492 & 21 & 12.8 & 0.0051 \\ 
PKS\,0859-14   & 2 & 3992 & 67 & 12.7 & 0.0032 \\
PKS\,1127-14   & 2 & 6268 & 81 & 12.8 & 0.0021 \\ 
PKS\,1136-13   & 2 & 6016 & 265 & 12.2 & 0.0020 \\
\\
PKS\,1144-379  & 2 & 1415 & 76 & 12.4 & 0.0088 \\ 
PKS\,1229-02   & 2 & 1927 & 121 & 13.1 & 0.0069 \\ 
PKS\,1245-19   & 2 & 6644 & 338 & 12.1 & 0.0013 \\  
PKS\,1508-05   & 2 & 5248 & 100 & 12.8 & 0.0024 \\ 
PKS\,1935-692  & 2 & 1641 & 39 & 16.2 & 0.0064 \\ 
\\
PKS\,2106-413  & 2 & 1790 & 13 & 12.4 & 0.0069 \\  
PKS\,2123-463  & 2 & 1521 & 46 & 11.3 & 0.0075 \\  
PKS\,2131-021  & 2 & 1957 & 14 & 11.1 & 0.0057 \\  
PKS\,2204-54   & 2 & 2356 & 20 & 11.9 & 0.0050 \\   
PKS\,2223-05   & 2 & 8356 & 359 & 21.9 & 0.0012 \\   
\\
PKS\,2244-37   & 2 & 1581 & 87 & 10.9 & 0.0069 \\   
\\
\multicolumn{5}{l}{Repeat observations}  \\
PKS\,1610-77   & 2 & 3584 & 134 & 12.7 & 0.0035 \\ 
MRC\,1908-201  & 2 & 1968 & 76 & 13.0 & 0.0066 
 \LL
}

\subsection{Quality of the ASKAP radio spectra}
\begin{figure*}
\centering
\includegraphics[width=0.9\textwidth]{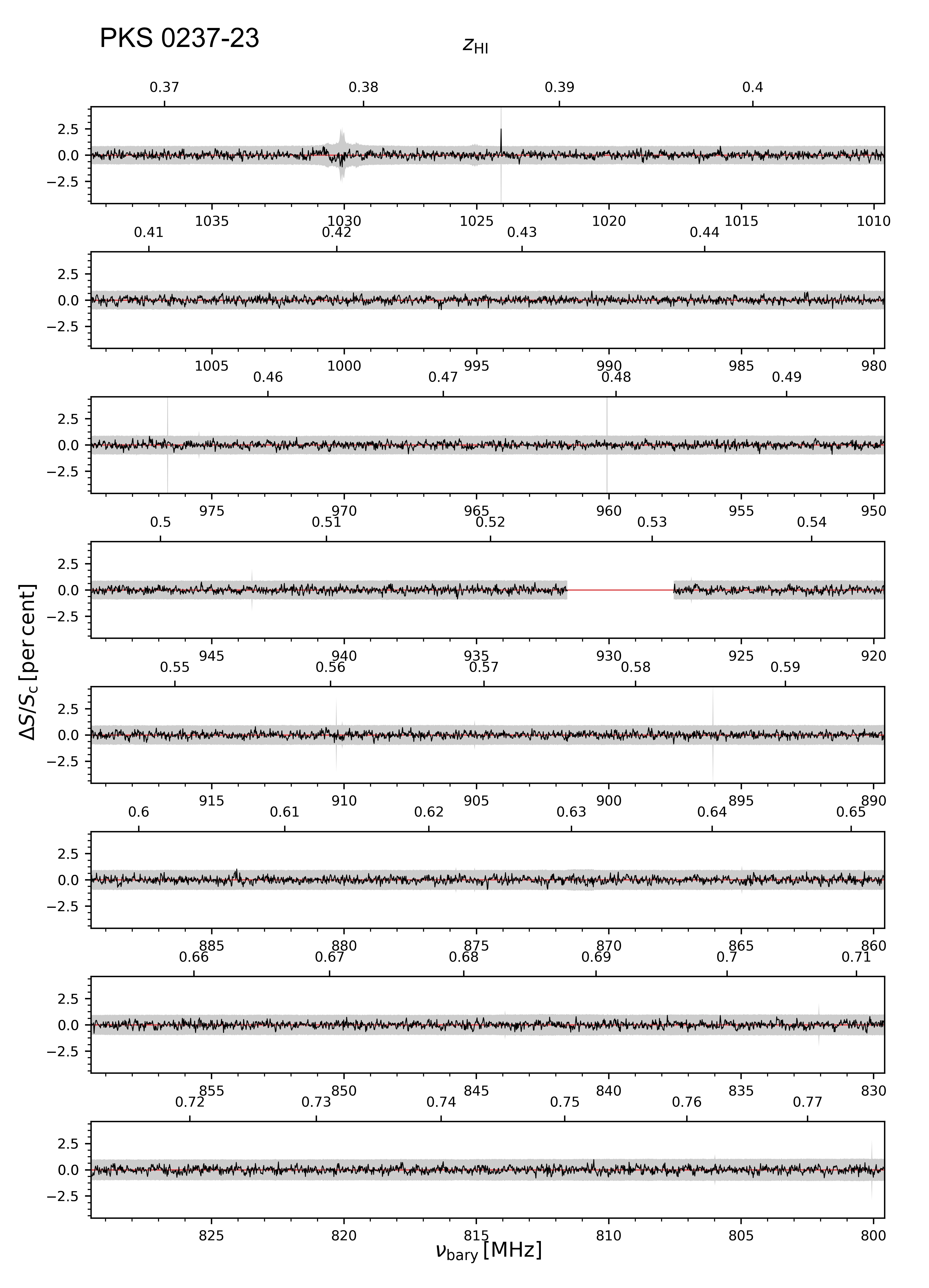}
\caption[]{The full ASKAP-12 spectrum of PKS\,0237-23 (AT20G J024008-230916), a radio-loud quasar at $z=2.223$. The median rms noise in this spectrum is 12.6\,mJy per 18\,kHz  channel.  
Some noise spikes are visible, and one small region (near 930\,MHz) is missing data because of a correlator-block failure. The light-grey band is set at $\pm$5 times the rms noise. }
\label{fig:full_spec}
\end{figure*}

Figure \ref{fig:full_spec} shows a full ASKAP-12 spectrum of one of our target sources, PKS\,0237-23. The quality of the spectrum is typical of those obtained during commissioning time, and the band is completely free of terrestrial and satellite-generated RFI. 

The rms noise level in this spectrum is roughly constant with frequency across the full 288\,MHz ASKAP band, giving a similar detection sensitivity at all redshifts sampled, and this is also true of the radio spectra of our other BETA and ASKAP-12 target sources. However, in the case of ASKAP-12 we found an error in the firmware weights that are used to correct the 1-MHz channelisation. This generates features at the level of 1\,per\,cent in gain amplitude every 1\,MHz, which we account for when assessing detection reliability.

\subsection{Detection limits and sensitivity}
\begin{figure*}
\centering
\includegraphics[width=0.49\textwidth]{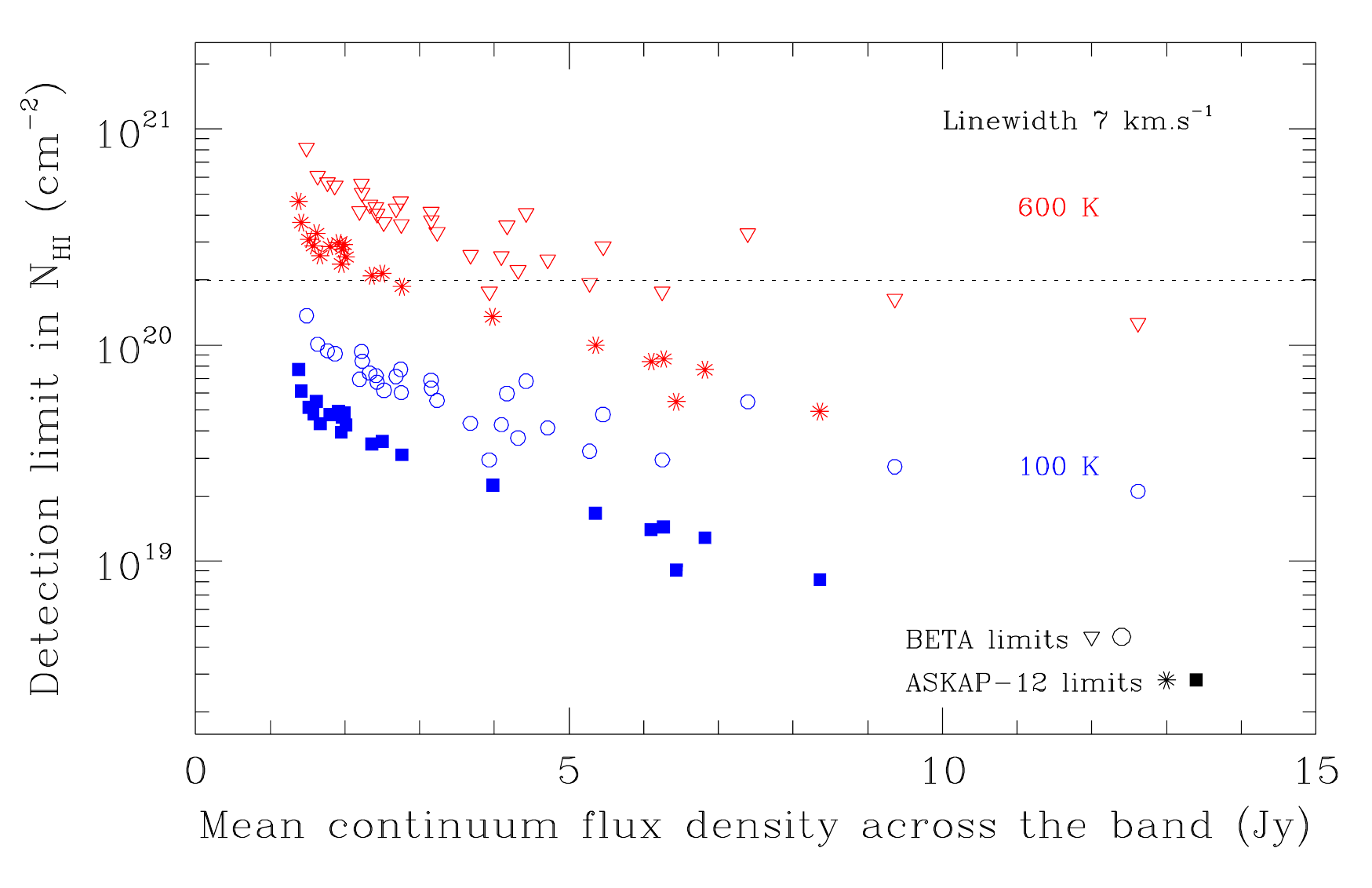}
\includegraphics[width=0.49\textwidth]{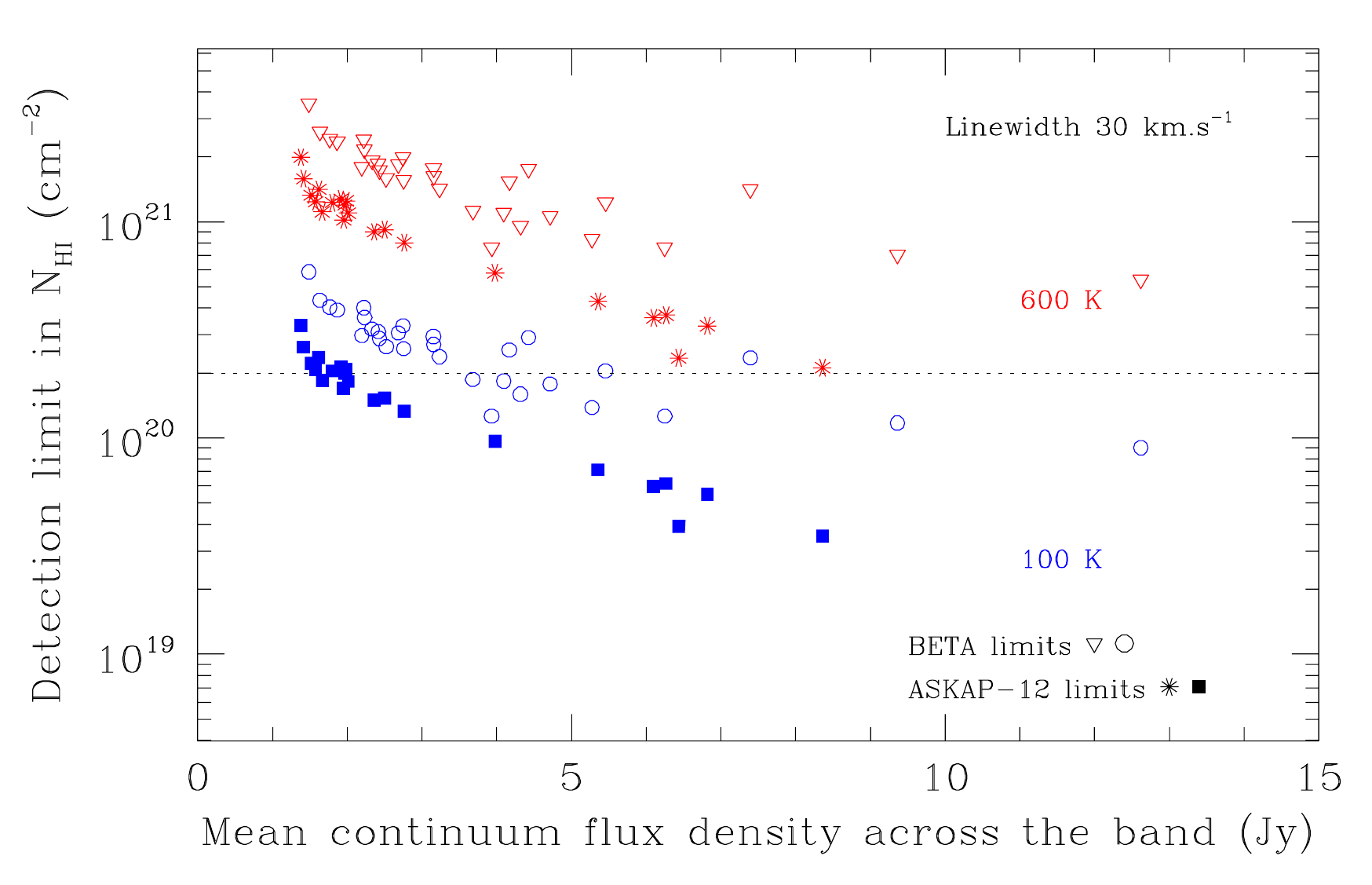}
\caption[]{Detection limits in HI column density for the observations in Tables \ref{tab:sample_data} and \ref{tab:sample_data2}, assuming a covering factor $f=1$, and a spin temperature of 100\,K (blue squares and open circles) or 600\,K (red stars and open triangles). The left-hand plot assumes a $5\sigma$\ detection limit for a narrow line of width 7\,km\,s$^{-1}$, while the right-hand plot shows a $5.5\sigma$\ detection limit for a broader line with a width of 30\,km\,s$^{-1}$ as discussed by Allison et al.\ (2020). The horizontal dotted line in each plot shows the DLA column density limit of $2\times10^{20}$\,cm$^{-2}$. }
\label{fig:phot3}
\end{figure*}

As can be seen from Tables 3 and 4, the rms noise in our spectra was typically 20--30 mJy per spectral channel for BETA observations and around 12 mJy per channel for observations with ASKAP-12. This corresponds to a 5$\sigma$ detection limit in optical depth of around 0.05 for BETA and 0.02 for ASKAP-12. 

Figure \ref{fig:phot3} plots the detection limits in HI column density for each of the spectra in Tables \ref{tab:sample_data} and \ref{tab:sample_data2}, assuming a covering factor $f=1$ and HI spin temperatures of 100\,K (blue points) and 600\,K (red points). For $T_{\rm s} = 100$\,K, we should be able to detect a minimal DLA absorber with $N_{\rm HI}\sim 2\times10^{20}$\,cm$^{-2}$ for all our targets across the full range of redshifts probed. The red points in Figure \ref{fig:phot3} are included to show that DLAs with spin temperatures above about 600\,K are unlikely to be detectable in the current pilot survey. 

\begin{figure*}
\centering
\includegraphics[width=0.48\textwidth]{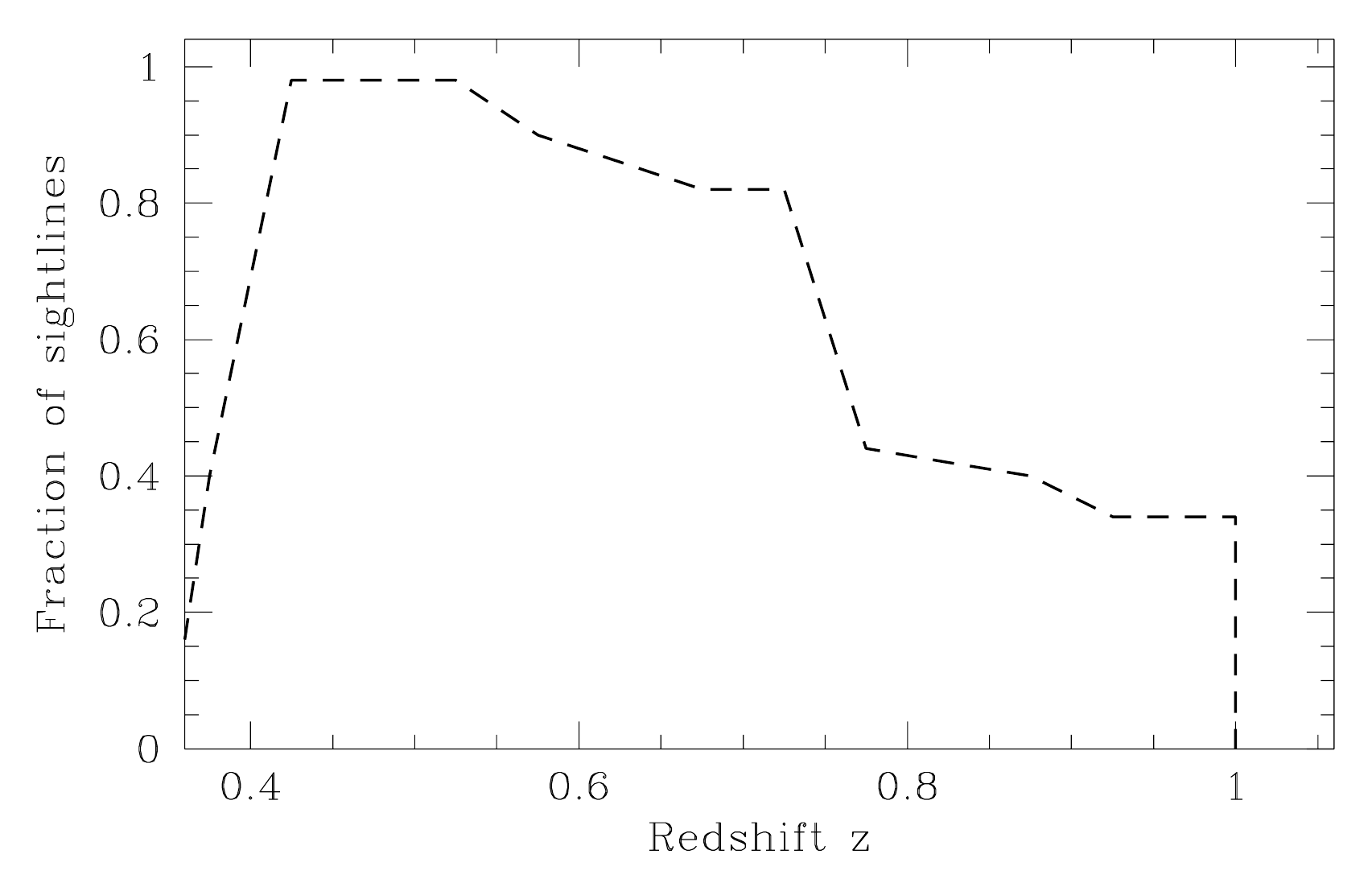}\includegraphics[width=0.48\textwidth]{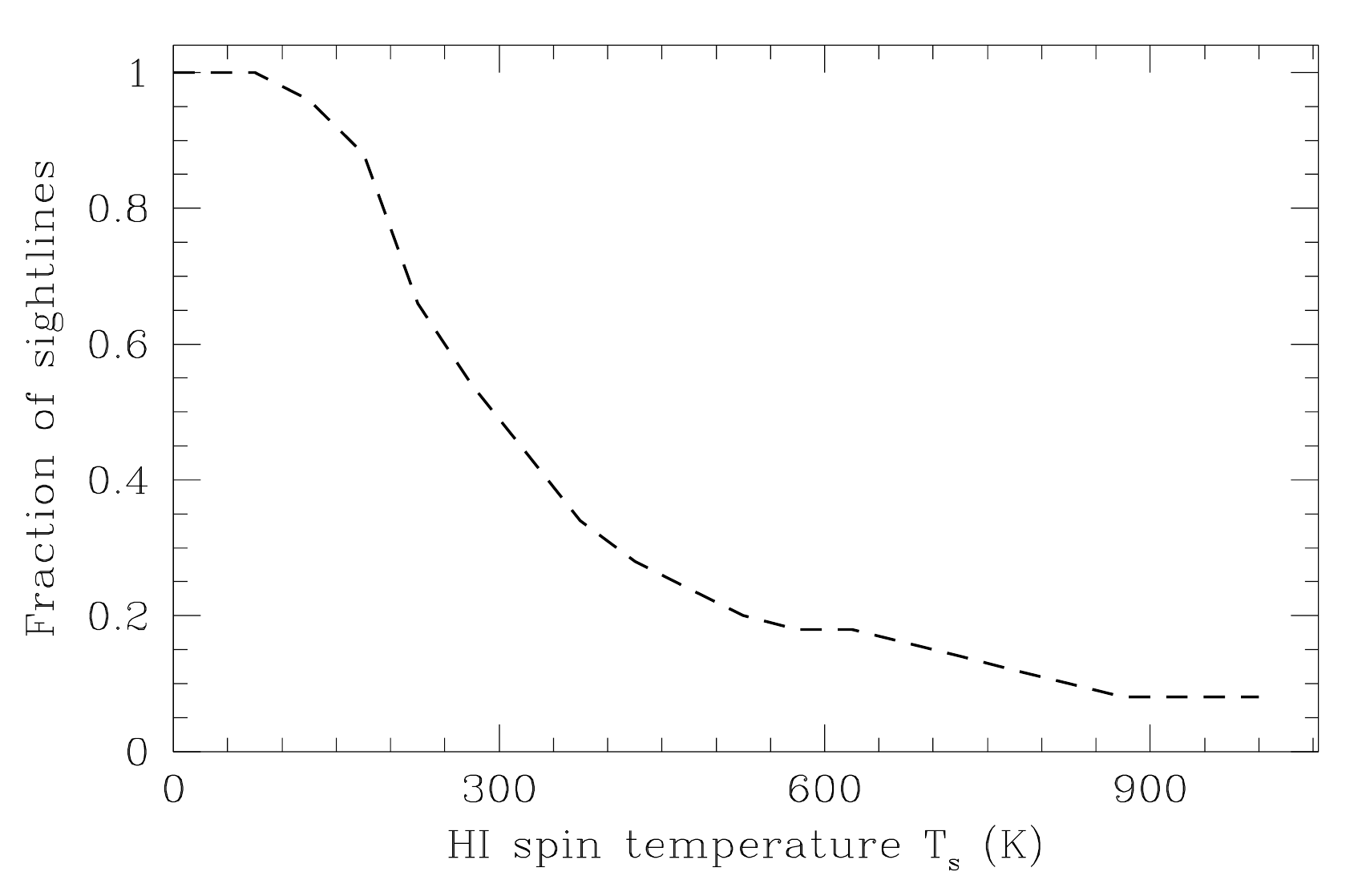}
\caption[]{(left) Fractional redshift coverage for the radio sources in Tables 1 and 2. This is set mainly by the frequency range used for the BETA and ASKAP-12 observations, so we probe fewer sightlines at $z>0.77$. (right) The fraction of sightlines for which we could detect (at the 5-$\sigma$ level) a minimal DLA system with $N_{\rm HI}=2\times10^{20}$\,cm$^{-2}$, assuming a line width 7\,km\,s$^{-1}$ and covering factor $f=1$.
For spin temperatures above 250-300\,K, detections of most DLA systems are only possible against the strongest continuum sources in our sample. }
\label{fig:dz}
\end{figure*}

Figure \ref{fig:dz} gives a more detailed look at the redshift and spin-temperature sensitivity of the sample as a whole. In these plots, the vertical axis shows the fraction of our sightlines on which we could detect (a) a sufficiently-strong intervening HI absorption line at redshift $z$, and (b) a minimal DLA system (with $N_{\rm HI}=2\times10^{20}$\,cm$^{-2}$ and covering factor $f=1$). This `fraction of sightlines' roughly corresponds to the total redshift path-length $\Delta z$ over which each kind of line could be detected in our survey. 
While this path-length is roughly constant with redshift (dropping off at $z>0.8$ because of the frequency limit of the ASKAP-12 spectra), the right-hand plot shows that (as expected for a 21\,cm survey) we are much more sensitive to absorption lines from gas with a low HI spin temperature than to lines that originate in warmer gas. Figure \ref{fig:dz} shows that that our 50\% completeness corresponds to $T_{\rm S}\sim300$\,K. We note that this is similar to the harmonic mean spin temperature of the Milky Way ISM \cite{murray18}.

\section{Results}\label{sec:results} 

\subsection{Detection of redshifted 21\,cm absorption lines}
The total path-length over which we could detect 21\,cm absorption lines was $\Delta z$ = 21.37 for the 50 background radio sources with known optical redshifts. We made no prior assumptions about the redshift of any line, so this is a genuinely `blind' line search in the spectral domain. 

We used the Bayesian detection method developed by \citealt{allison12} to search for absorption lines in the spectrum. This method uses the Bayes factor to assign significance to a particular feature in the spectrum, equal to ratio of the Bayesian evidences for a Gaussian line and noise model, versus a null model comprising just the noise. We consider all features with values greater than $\ln{\mathrm{Bayes factor}} > 1$. Given the aforementioned error from applying incorrect coarse-channelisation weights, and failures of individual correlator cards, further visual inspection of detected features was required. Upon inspection, we obtained four reliable detections of intervening absorption lines and one associated absorption line along the line of sight to the 53 radio sources listed in Tables 1 and 2.

The associated absorption-line system, at $z=0.44$\ in PKS\,1740-517, has been studied in detail by \cite{allison15} and \cite{allison19} and is not discussed further in this paper.

The properties of the four intervening absorption lines are summarised in Table \ref{tab:beta_det} and discussed below. Two of these lines (towards PKS\,0834-20 and PKS\,1610-77) are new detections, while the other two are re-detections of previously-known absorption-line systems. 

\subsection{Intervening HI absorption towards PKS 0834-20 }
\begin{figure*}
\centering

\vspace*{0.2cm}
\includegraphics[width=0.48\textwidth]{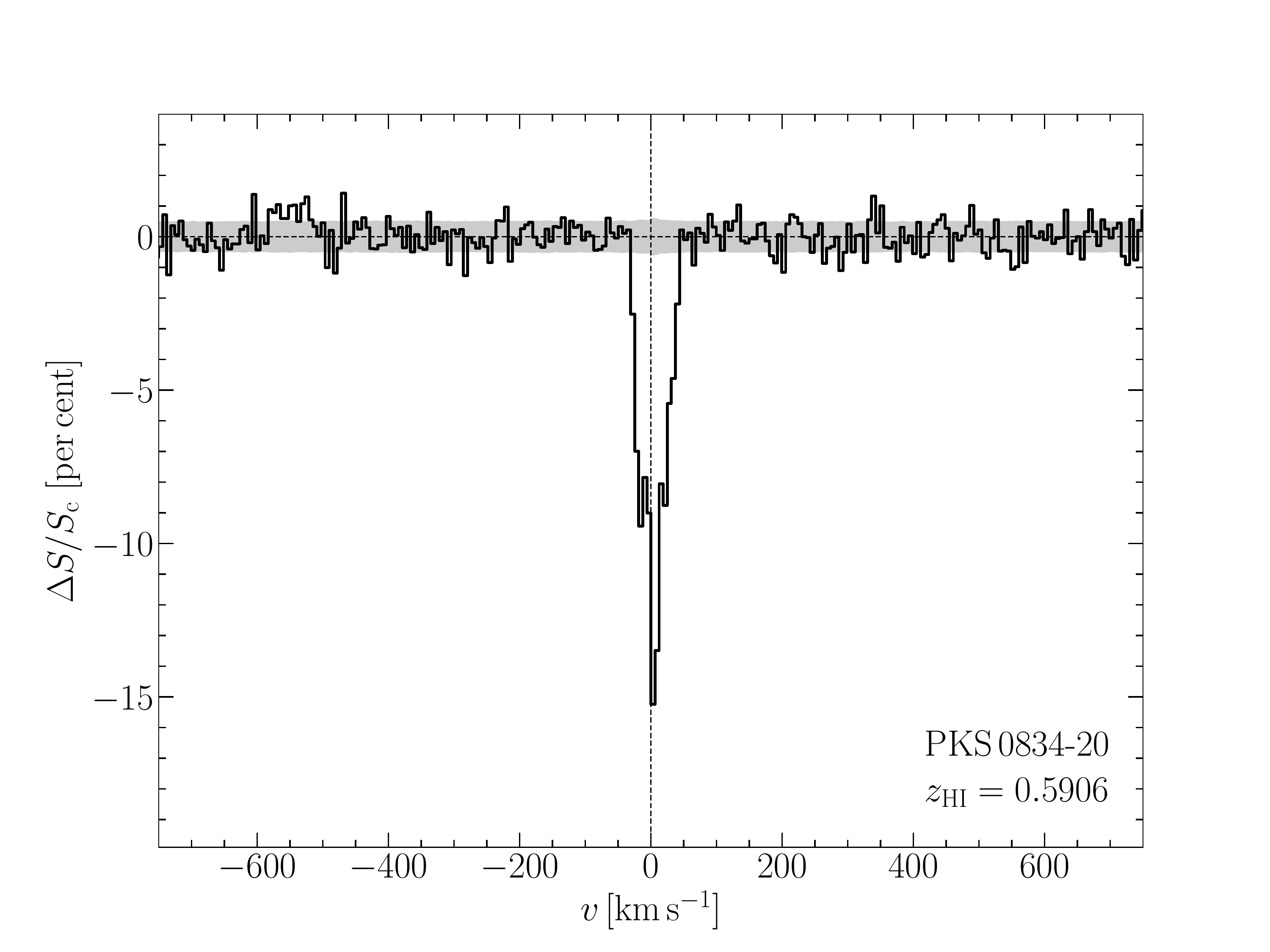}
\includegraphics[width=0.475\textwidth]{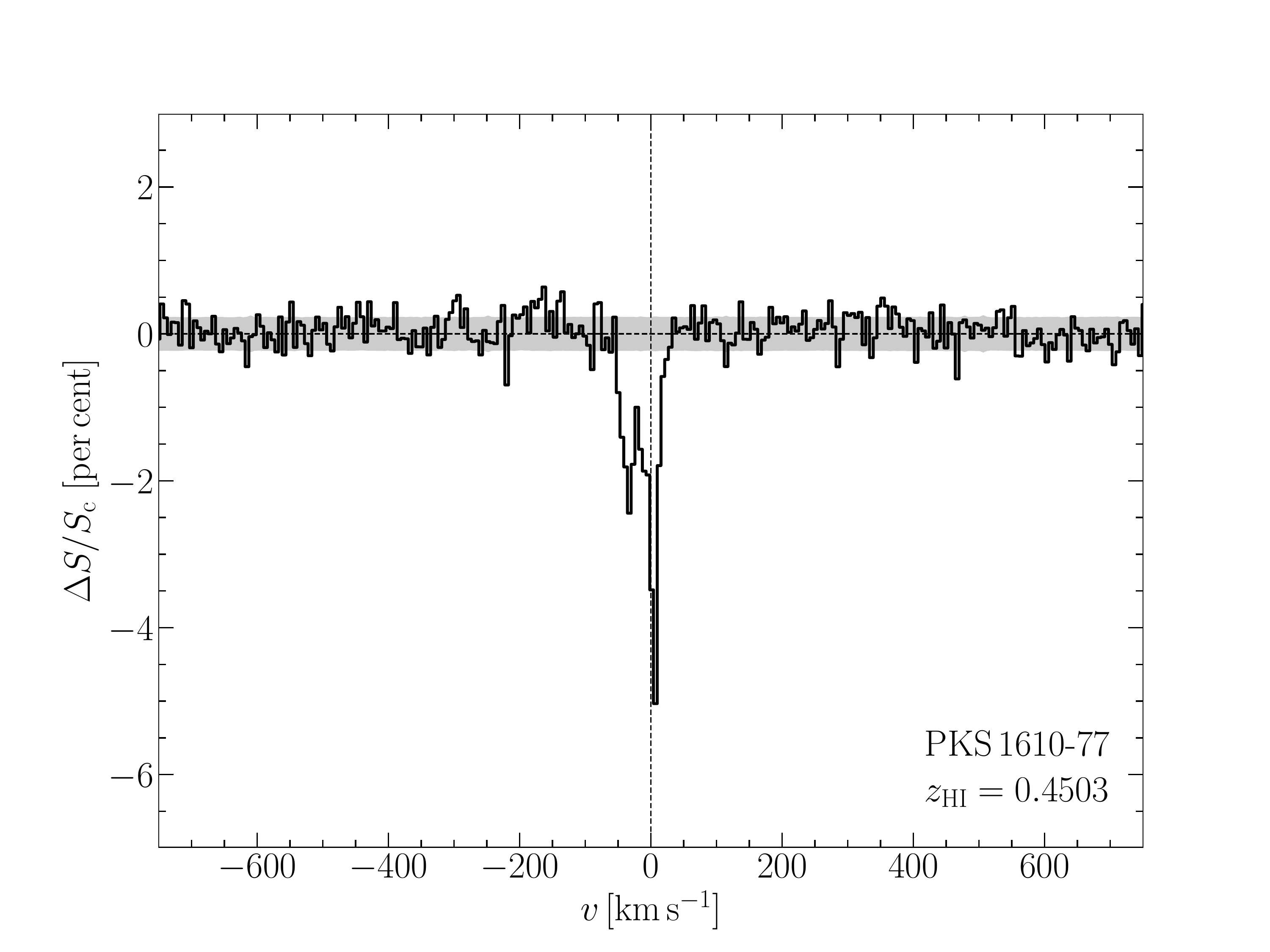}

\vspace*{0.5cm}
\includegraphics[width=0.47\textwidth]{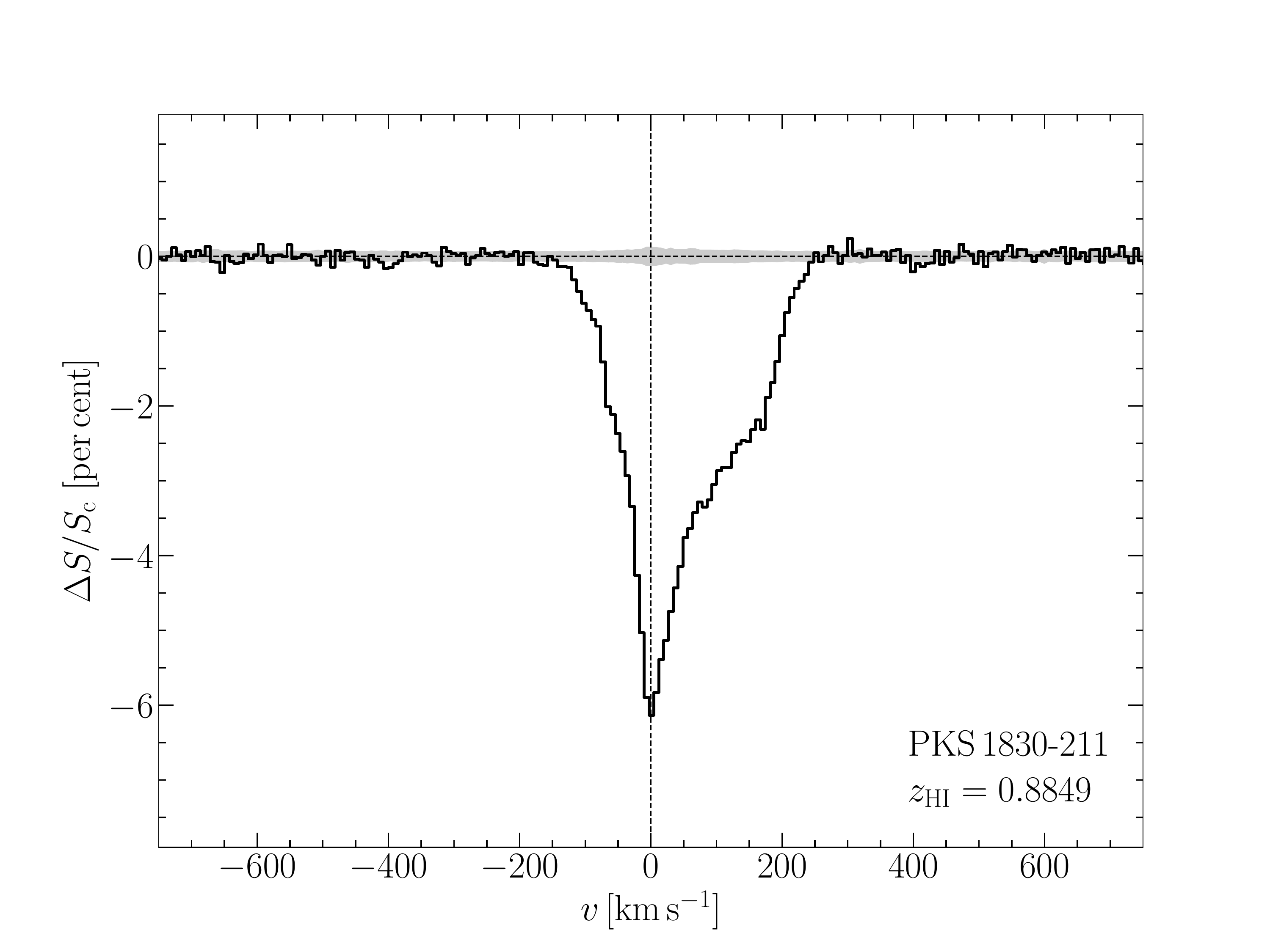}
\caption[]{Intervening absorption lines detected by ASKAP towards the radio sources PKS\,0834-20, PKS\,1610-77 and  PKS\,1830-211.  
The grey shading in each plot shows $\pm\tau_{\rm lim}$, where $\tau_{\rm lim}$ is the 1$\sigma$ limit in optical depth listed in Tables 4 and 5. }
\label{fig:abs_new}
\end{figure*}

We made a new detection of 21\,cm HI absorption at $z=0.591$ along the line of sight to the background ($z=2.75$) radio source PKS\,0834-20. Figure \ref{fig:abs_new} shows the ASKAP spectrum at the position of the absorption line. The peak optical depth of this line ($\tau=0.14$) is the highest out of our four ASKAP detections. 

The background radio source, PKS\,0834-20 is a radio-loud blazar, and low-frequency observations show a broad peak in the radio spectrum near 500\,MHz  \citep{callingham17}.  
The 15\,GHz VLBA continuum image published by the MOJAVE team \citep{pushkarev17} shows a core-jet structure with a total extent of 6.4\,mas, corresponding to 42\,pc at the redshift of the 21\,cm absorption line. 

\begin{figure*}
\centering
\includegraphics[width=0.48\textwidth]{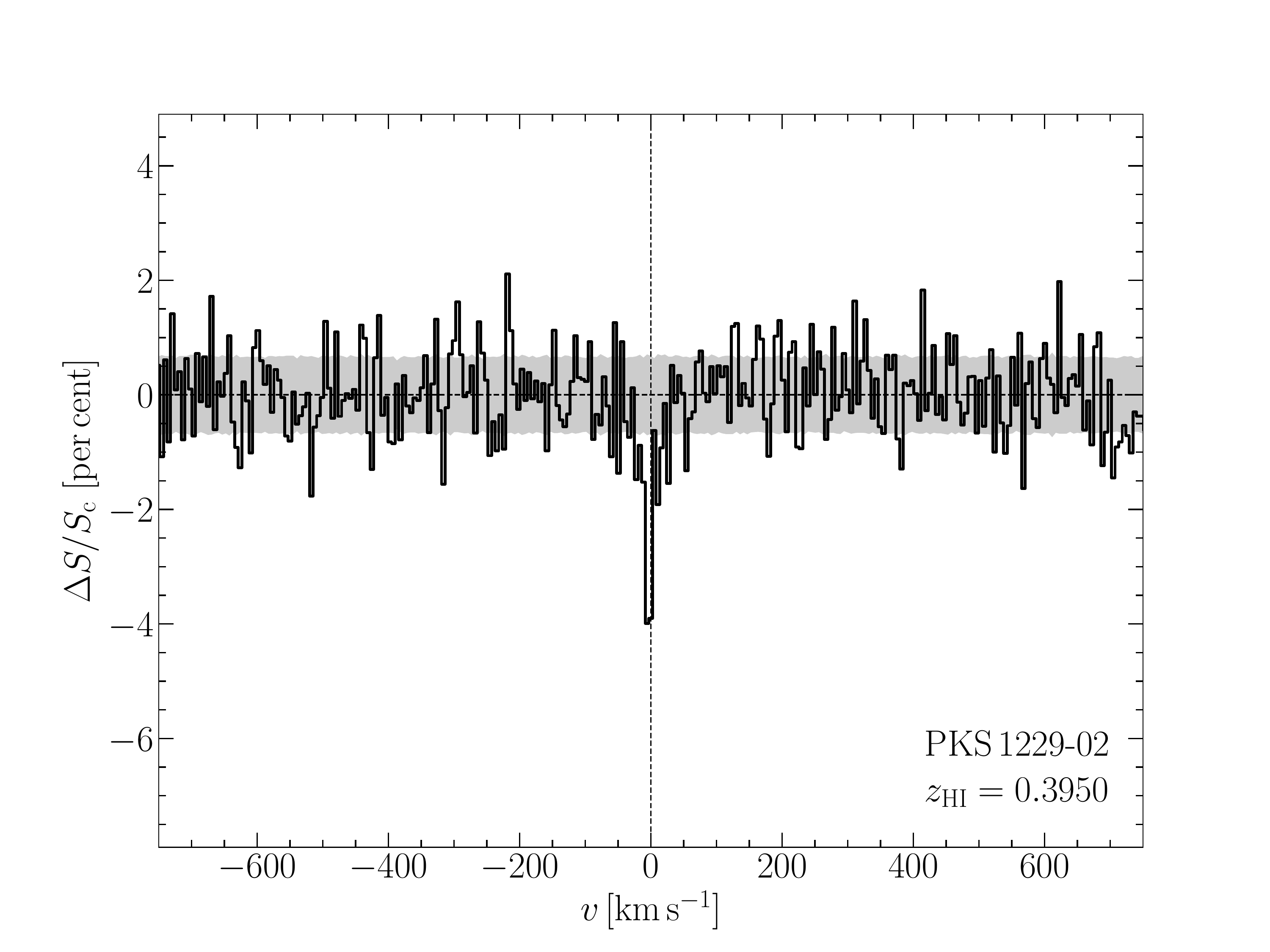}
\includegraphics[width=0.48\textwidth]{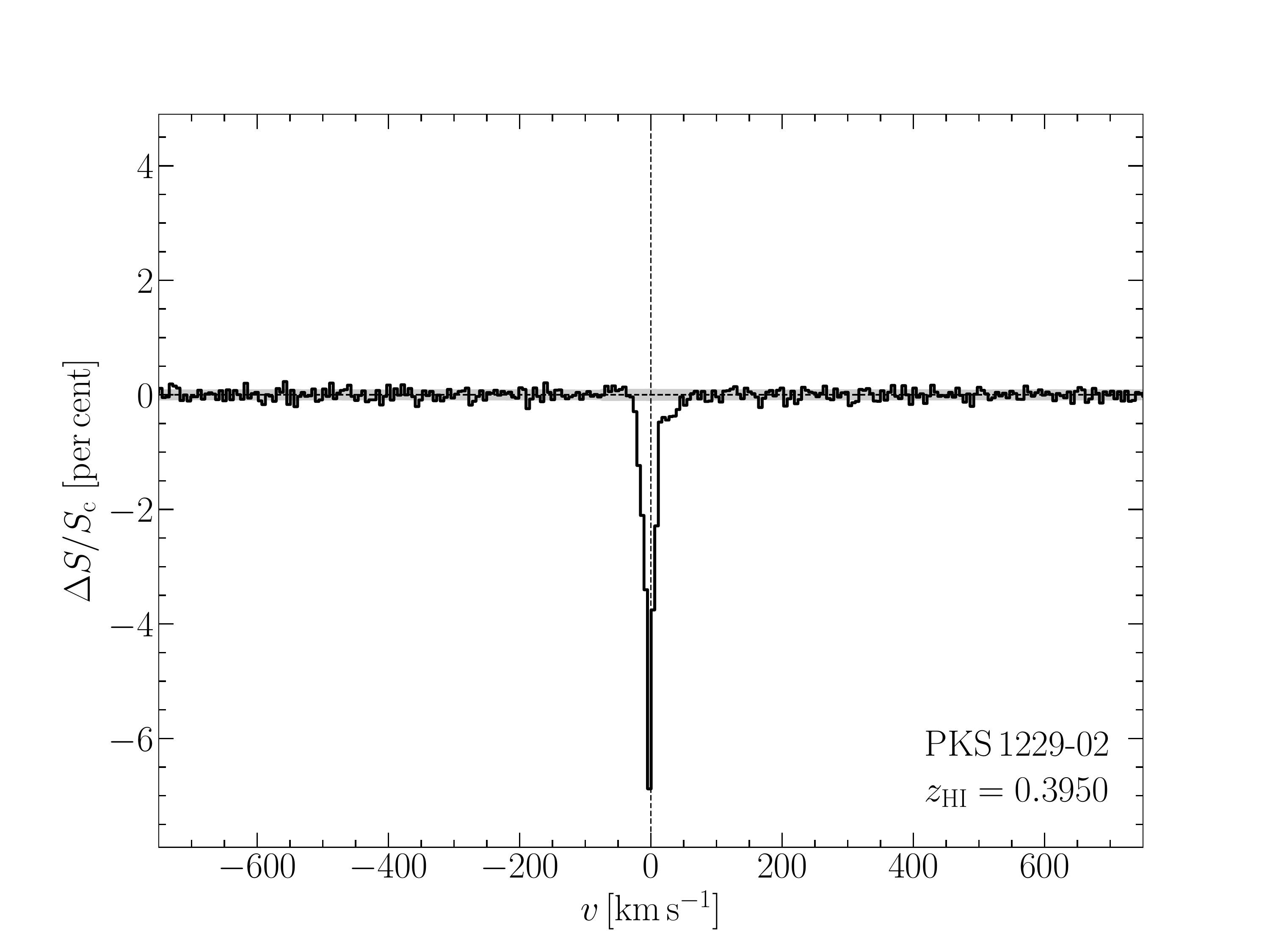}
\caption[]{Intervening absorption at $z=0.3950$ along the line of sight to the background radio source PKS\,1229-02 at $z=1.045$. \\ (left) The original ASKAP-12 spectrum, (right) A new ASKAP-36 spectrum of the same sightline. 
The grey shading shows $\pm\tau_{\rm lim}$, where $\tau_{\rm lim}$ is the 1$\sigma$ limit in optical depth listed in Tables 4 and 5. }
\label{fig:det_1229}
\end{figure*}
  
\subsection{Intervening absorption towards PKS\,1229-02} 
\label{sec:1229}
A 21\,cm HI absorption line at $z=0.395$ was first detected along a sightline to PKS\,1229-02 by \cite{brown79} with the NRAO 140-ft (43\,m) single-dish radio telescope, and this observation was motivated by the presence of strong Mg\,II absorption in the optical spectrum \citep{kinman67}. \cite{Wolfe80} detected the same HI absorption line with the Arecibo telescope, again using MgII pre-selection. 

Figure \ref{fig:det_1229} shows the ASKAP detection of intervening 21\,cm absorption at $z=0.395$ along the line of sight to the radio-loud QSO PKS\,1229-02.  
The original ASKAP-12 detection had relatively low S/N, and the measured optical depth of the HI line was significantly lower than the values observed by \cite{brown79} and \cite{Wolfe80}. To test whether this was the result of variability in the line profile over time, we re-observed PKS\,1229-02 in a test observation for 8 hours with the full 36-antenna ASKAP array in May 2019. The new ASKAP-36 spectrum is also shown in Figure \ref{fig:det_1229}. The total integration time for the ASKAP-36 spectrum was 8 hours, and the HI optical depth measured from this higher-quality ASKAP spectrum is similar to the \cite{brown79} and \cite{Wolfe80} values. 

The PKS\,1229-02 absorption system is well-studied at optical and UV wavelengths. The strong metal absorption lines in the optical spectrum at $z=0.395$ have been studied by \cite{briggs85} and \cite{lanzetta92}, and imply a relatively high metallicity for the absorbing gas. 

The PKS\,1229-02 absorption system is the only one in our current sample for which an HI column density has also been measured from the Ly$\alpha$ line. 
\cite{boisse98} used the HST to observe the damped Lyman-$\alpha$ line in the UV at $z=0.395$, from which they measured an HI column density of $N_{\rm HI}$ = $5.6\times10^{20}$\,cm$^{-2}$. 
In principle, this allows us to measure the HI spin temperature $T_{\rm s}$ if we assume that the Ly$\alpha$ and 21\,cm absorption measurements are along the same sightline. 
Unlike most of the other sources in our sample, PKS\,1229-02 contains extended radio structure on scales out to $\sim18$\,arcsec \citep{hintzen83,kronberg92} in addition to a compact central source. This complicates the analysis of the absorption-line spectrum, and we adopt the covering factor of $f=0.42$ derived by \cite{kanekar09}. 

From our ASKAP spectrum, we derive a value of  $T_{\rm s} =102\pm12$\,K for the PKS\,1229-02 absorption system.   This is consistent with the previously-published estimates of $T_{\rm s}$ summarized in Table \ref{tab:spin}, which range from 95 to 170\,K. 

\subsection{Intervening absorption towards PKS 1610-77} 
With BETA, we made a new detection of intervening 21\,cm absorption at $z=0.4503$ along the line of sight to the radio-loud QSO PKS\,1610-77, as shown in Figure \ref{fig:abs_new}. The line has at least two velocity components, with a separation of $\approx 30$\,km s$^{-1}$. 

The VLBI image published by \cite{ojha10} shows a curved jet extending about 5\,mas from the core, along with some diffuse emission on larger scales. At the redshift of the HI absorption line, the angular extent of this jet corresponds to a linear size of $\sim30$\,pc. 

\subsection{Intervening absorption towards PKS\,1830-211} 
PKS\,1830-211 is a gravitationally-lensed radio source \citep{subrahmanyan90, jauncey91} with an intervening galaxy at $z=0.886$ \citep{wiklind96} and a possible second galaxy at $z = 0.192$ \citep{lovell96}. 

\cite{chengalur99} first detected HI (and OH) absorption at $z = 0.886$ in this system using the Westerbork Synthesis Radio Telescope. 
\cite{allison17} obtained ASKAP spectra of PKS\,1830-211 from 700-1530\,MHz, re-detecting HI absorption at $z = 0.192$ and $z = 0.886$, and OH absorption at $z = 0.886$. Comparing spectra for several epochs spanning 20 years, they identified variability in the HI line, consistent with changes in the background quasar. We include the $z = 0.886$ HI line detected with ASKAP in our sample here. 

\ctable[
notespar,
star,
cap = {BETA obs},
caption = {Parameters of the detected 21\,cm absorption lines. The estimated values of $N_{\rm HI}$\ assume $T_{\rm s}=100$\,K and $f=1$ (col 6), and equation 9 of Braun 2012 (col 7). Since there can be multiple velocity components in a line, an effective width was determined by dividing the integrated optical depth in this table by the peak value \protect\cite[see e.g.][]{dickey82,allison13,allison14} - these effective velocity widths are listed in column 8. 
}, 
label = {tab:beta_det}
]{llcllrrrr}%
{ 
}{
\FL
\multicolumn{1}{c}{Name} & \multicolumn{1}{c}{$z_{\rm abs}$} & \multicolumn{1}{c}{$N_{\rm Gauss}$} & \multicolumn{1}{c}{$\tau_{\rm pk}$} & \multicolumn{1}{c}{$\int \tau\ {dV}$} &\multicolumn{2}{c}{Estimated HI column} &  \multicolumn{1}{c}{Effective} & \multicolumn{1}{c}{Notes} \\
 & &  & &   \multicolumn{1}{c}{(km.s$^{-1}$)} &  \multicolumn{2}{c}{ density $N_{\rm HI}$ (cm$^{-2}$)} & \multicolumn{1}{c}{width} \\  
  &&&&& \multicolumn{1}{c}{[$T_{\rm s}$=100\,K]} & 
  [Braun2012] & \multicolumn{1}{c}{(km.s$^{-1}$)} & \\ 
\multicolumn{1}{c}{(1)} & \multicolumn{1}{c}{(2)} & \multicolumn{1}{c}{(3)} & \multicolumn{1}{c}{(4)} &
\multicolumn{1}{c}{(5)} &
\multicolumn{1}{c}{(6)} & \multicolumn{1}{c}{(7)} & \multicolumn{1}{c}{(8)} & \multicolumn{1}{c}{(9)} \\
\hline \hline 
 PKS\,0834-20  & 0.5906 & 3 & $0.1596\pm0.0051$ & $6.178\pm0.112$ & $1.1\times10^{21}$ & 
  $1.2\times10^{21}$ & 38.7 & ASKAP-12 \\
 PKS\,1229-02 & 0.3950 & 3 & $0.0699\pm0.0012$ & $1.233\pm0.021$ & $2.2\times10^{20}$ & 
 $5.3\times10^{20}$ & 17.6 & ASKAP-12/36 \\
 PKS\,1610-77  & 0.4503 & 2 & $0.0491\pm0.0027$ & $1.502\pm0.058$  &  $2.7\times10^{20}$ & 
 $3.7\times10^{20}$ & 30.6 & BETA \\ 
 PKS\,1830-211 & 0.8849 & 3 & $0.0629\pm0.0008$ &  $9.792\pm0.050$ & $1.8\times10^{21}$ & 
 $4.8\times10^{20}$ & 155.7 & BETA 
 \LL}

\section{Discussion} 

\subsection{HI column density estimates}
As noted earlier (\S1.3), we need to know both the harmonic mean spin temperature $T_{\rm s}$ and the covering factor $f$ to convert the observed optical depth of a 21\,cm HI absorption line to an HI column density $N_{\rm HI}$. 

Even without accurate measurements of $T_{\rm s}$ and $f$ for individual sources, however, we can make some general statements about the likely HI column densities along the sightlines where we detected intervening 21\,cm absorption lines with ASKAP. 
In particular, we can consider how many of these detected lines are likely to arise in gas with an HI column density above the DLA threshold of $2\times10^{20}$\,cm$^{-2}$. This in turn will allow us to estimate the DLA number density n($z$) in the redshift range probed by our ASKAP spectra. 

Since most of the background radio sources in our survey are compact on scales of 100\,mas or less (as discussed in \S2.3), we will assume for now that the covering factor $f\sim1$. 
This is a conservative assumption in terms of measuring the total number of DLA systems, since if $f <1$, then the actual HI column density will be higher than that estimated from the observed 21\,cm optical depth. 
We note that the absorber in PKS\,1229-02 is likely to have a covering factor $f<0.5$ as discussed in section \ref{sec:1229}.  

\begin{figure}
\centering
\includegraphics[width=0.5\textwidth]{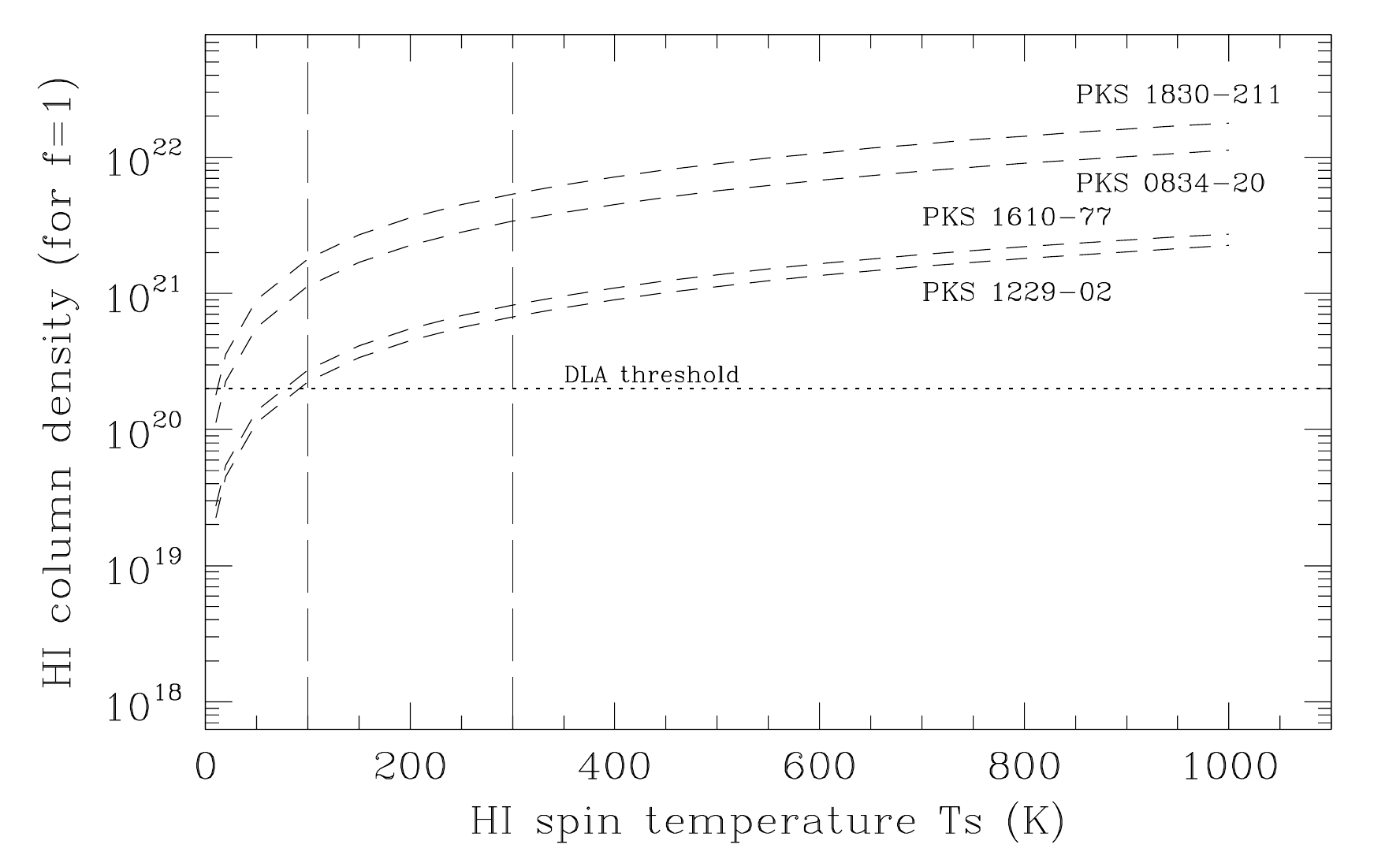}
\caption[]{HI column densities as a function of assumed spin temperature for the four detected absorption lines presented in \S4, derived from the observed optical depth of the line assuming a covering factor $f=1$. Vertical dashed lines at 100\,K and 300\,K show the approximate range of spin temperatures expected for the cold HI gas detected in absorption with ASKAP (based on the sensitivity plot shown in Figure 3). }
\label{fig:tspin}
\end{figure}

Figure \ref{fig:tspin} shows how the HI column density $N_{\rm HI}$ derived for our four detected lines varies with HI spin temperature. The values plotted assume a covering factor $f=1$, so the derived values of $N_{\rm HI}$ will be proportionally higher if $f<1$. 

We estimate $N_{\rm HI}$ from our detected absorption lines in two different ways:   
\begin{enumerate}
\item Taking the commonly-used fiducial value of $T_{\rm s}=100$\,K, which assumes all of the absorbing gas is in the CNM (\citealt{wolfire03}) and so provides a lower limit. 
\item Using the empirical relation between $\tau$ and $N_{\rm HI}$ derived by \cite{braun12}. 
\end{enumerate}

The values of $N_{\rm HI}$ derived under these two assumptions are listed in columns 6 and 7 of Table \ref{tab:beta_det}.  

\subsection{The 21\,cm DLA number density at $z\sim0.6$}
In terms of the number of DLA-like absorbers detected in our survey, the results in Table \ref{tab:beta_det} and Figure \ref{fig:tspin} appear reasonably consistent. Three absorbers (PKS\,0834-20, 1610-77 and 1830-211) all have HI column densities above the DLA threshold range both for a range of plausible values of $T_{\rm s}$ and $f$ and for the empirical \cite{braun12} relation. 

A fourth system, PKS\,1229-02, has a 21\,cm value closer to the DLA threshold if we assume $f\sim1$, but in this case 
we also have a direct HST Lyman-$\alpha$ measurement for PKS\,1229-02 of $N_{\rm HI}$ = $5.6\times10^{20}$\,cm$^{-2}$ \citep{boisse98} which confirms this as a DLA system (see \S \ref{sec:1229}). 

We therefore have four 21\,cm DLA detections on sightlines covering a total redshift interval $\Delta z = 21.37$. This yields an estimated DLA number density at redshift $z\sim0.6$ of dN/d$z$ = $0.19\substack{+0.15 \\ -0.09}$, where the quoted uncertainties correspond to $1\sigma$\ Gaussian errors calculated for small event numbers \citep{gehrels1986}. As can be seen from Figure \ref{fig:dz}, we are mainly sensitive to gas with a spin temperature typical of the CNM ($T_{\rm s} \lesssim 300$\,K) and would be unlikely to detect DLA systems arising in warmer gas. 
 
\begin{figure*}
\centering
\includegraphics[width=0.85\textwidth]{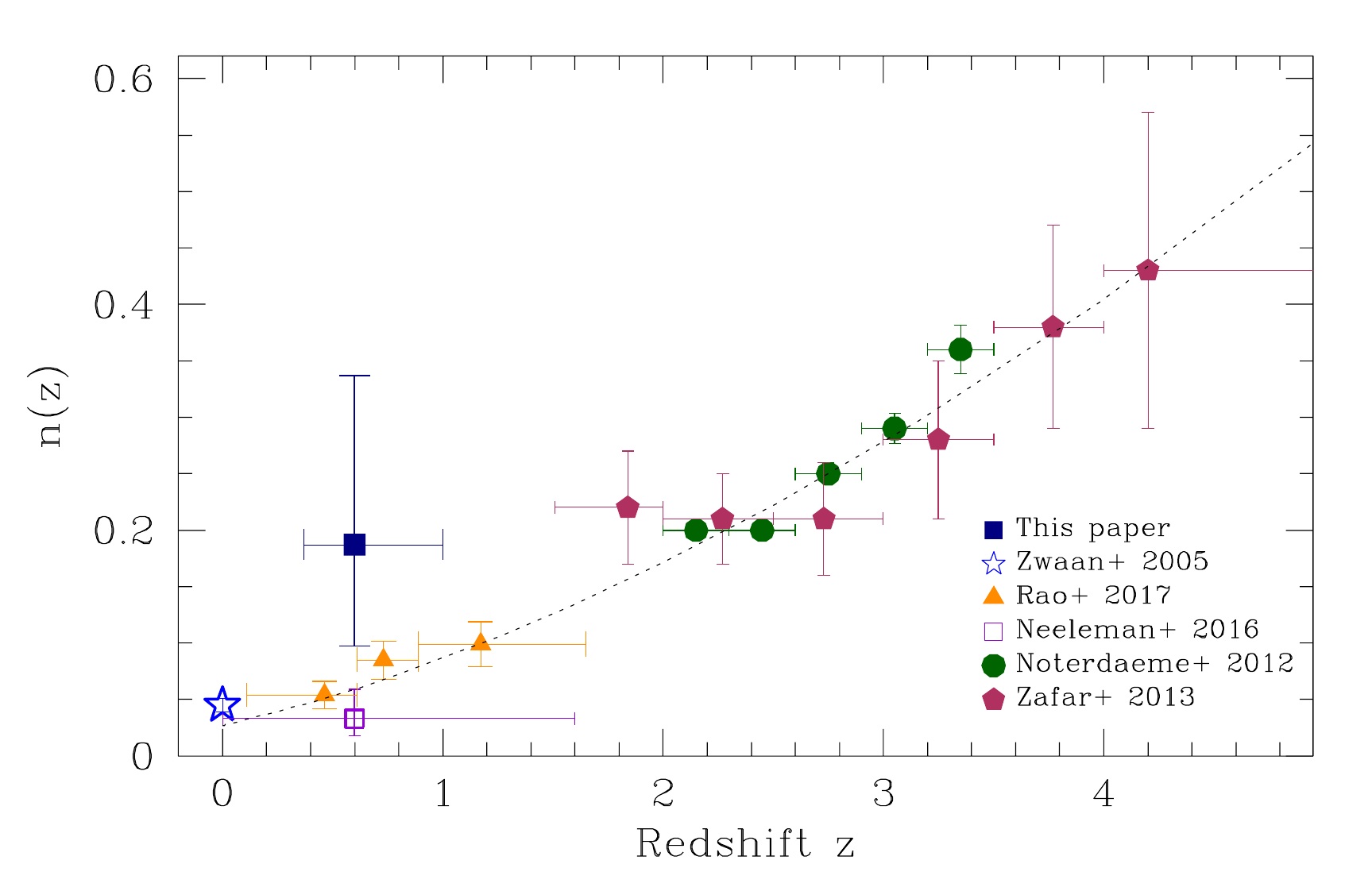}
\caption[]{The number density of DLA absorbers, n($z$),  as a function of redshift. The dark blue square at $z=0.6$ shows the 21\,cm value derived from our ASKAP pilot survey, as discussed in \S5.2 of this paper. The other values plotted are the 21\,cm HI emission point from Zwaan et al.\ (2005), HST DLA values from Neeleman et al.\ (2016) and Rao et al.\ (2017), and ground-based DLA measurements from Noterdaeme et al.\ (2012) and Zafar et al.\ (2013). The dashed line shows the empirical n($z$) vs $z$ relation from Rao et al.\ (2017).  }
\label{fig:ndla}
\end{figure*}

Figure \ref{fig:ndla} compares our result with those of recent optical DLA studies. The only other 21\,cm point is the local ($z\sim0$) value from \cite{zwaan2005}. The \cite{neeleman16} and \cite{rao17} values are both from HST observations of damped Ly$\alpha$ lines - the \cite{rao17} study preselected targets that showed MgII absorption, while \cite{neeleman16} used a smaller HST sample without MgII preselection. The \cite{noterdaeme12} study at $z>1.7$ used a large ground-based sample of over 6000 SDSS quasars with DLA detections, while \cite{zafar13} analyzed a sample of 122 quasar spectra  spanning the redshift range $1.5<z<5$. 

It is notable that the ASKAP n($z$) point at $z\sim0.6$ lies above the \cite{neeleman16} and \cite{rao17} values measured at similar redshift, though at this stage the large error bars on the ASKAP value mean that our result is also (just) consistent with the general trend seen in optical DLA studies. 
Our high value for n($z$) is somewhat surprising, since we are only sensitive to the subset of DLAs with low $T_{\rm s}$ - and so our n($z$) value\ might be expected to be a lower limit to the total value. As can be seen from Figure \ref{fig:tspin}, the HI column densities for our detected absorbers could only lie below the DLA limit if the covering factor $f = 1$  and the spin temperature dropped below 50--80\,K. This appears unlikely, due to the known properties of the CNM and the paucity of observed systems with such low spin temperatures. 
Figure \ref{fig:ndla} therefore suggests a potentially significant discrepancy with some of the QSO DLA studies,  possibly because we are picking up dusty systems that might not have been included in optical QSO surveys. 
A larger sample of 21\,cm HI absorption detections is needed to explore this question further. 

\ctable[  
notespar,
  star,
  cap = {Spin temperature estimates for PKS\,1229-02},
  caption={Spin temperature estimates for the $z=0.395$ absorption system towards PKS\,1229-02, using the Ly$\alpha$ $N_{\rm HI}$ value of $5.6\times10^{20}$\,cm$^{-2}$ from \cite{boisse98}},  
label={tab:spin},
  ]{l l ll llll l ll cc} 
{\tnote[]{  } }
{\FL 
\multicolumn{1}{l}{$T_{\rm s}$ (K)}  & \multicolumn{1}{l}{Author} & \multicolumn{1}{l}{Notes} \\
 \hline\hline
 170 & \cite{boisse98} & Used \cite{brown79} HI data, which assumes $f=0.5$ & \\    
170 & \cite{chengalur00} & Recomputed from \cite{brown79} and \cite{briggs99} HI data & \\
105$\ \pm 30$ & \cite{lane2000} & Used \cite{briggs99} HI data \\
\ 95$\ \pm15$ & \cite{kanekar14a} & Used \cite{brown79} HI data and $f=0.42$\\
102$\ \pm12$ & This paper & Using ASKAP HI data, and assuming $f=0.42$ 
\LL }

\subsection{The nature of the intervening galaxies} 
Since our 21\,cm intervening absorbers were selected without any optical pre-selection, identifying their host galaxies gives us a first look at the {\it kinds}\ of galaxies that are present in an HI-selected sample in the distant Universe.  
In contrast to the host galaxies of the high-redshift ($z>1.7$) DLA systems detected in ground-based optical QSO surveys, the lower-redshift host galaxies of our 21\,cm absorbers may be bright enough to detect in high-quality  optical images even in the presence of a bright nearby quasar. For example, a typical spiral galaxy at $z\sim0.5$ is expected to have an I-band magnitude of 20-22 \cite{cantale16} and should be visible in good-quality images from an 8\,m-class telescope. 

Table \ref{tab:hosts} summarises the information currently available for the host galaxies of the four HI absorption systems detected in this study, and we discuss each of these absorption systems in turn below. 

\begin{figure}
\centering
\includegraphics[width=0.46\textwidth]{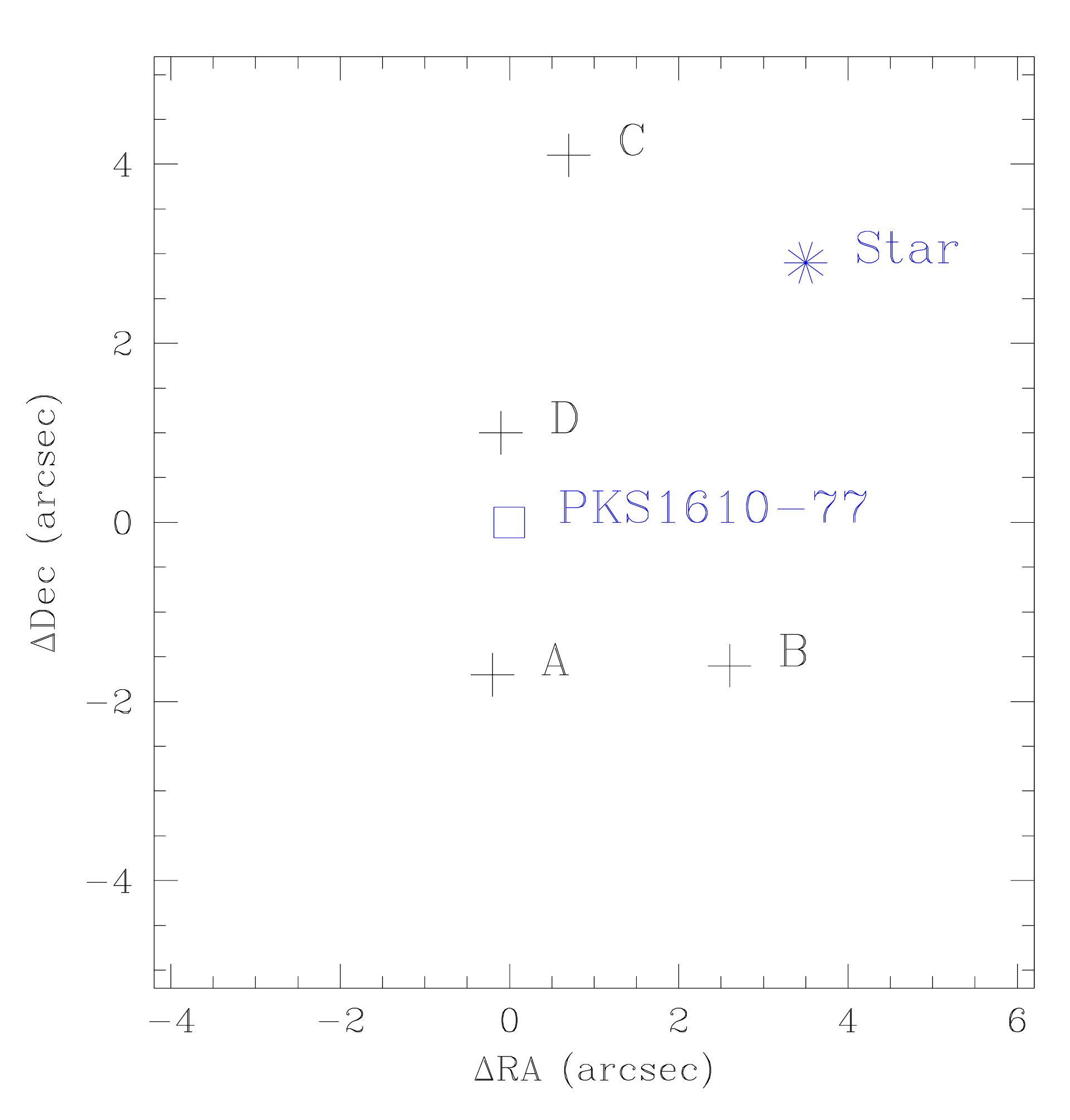}
\caption[]{Schematic representation of the positions of four galaxies (A, B, C, D) along the line of sight to the radio source PKS\,1610-77, based on data from Table 1 of \cite{courbin97}. The positions of the radio source and a foreground Galactic star are also shown. }
\label{fig:1610_gal2}
\end{figure}

\begin{figure*}
\centering
\includegraphics[width=0.8\textwidth]{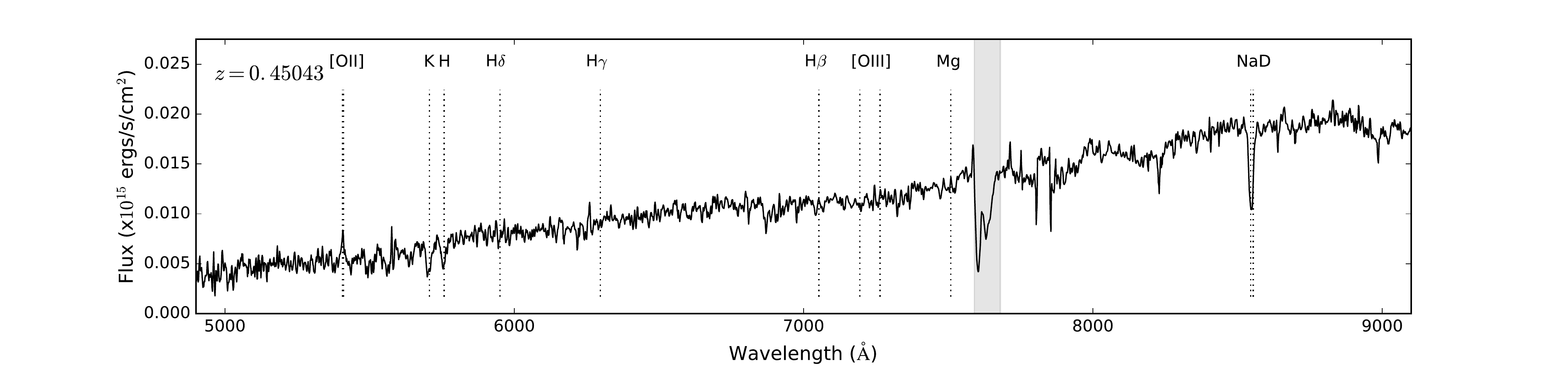} 
\includegraphics[width=0.8\textwidth]{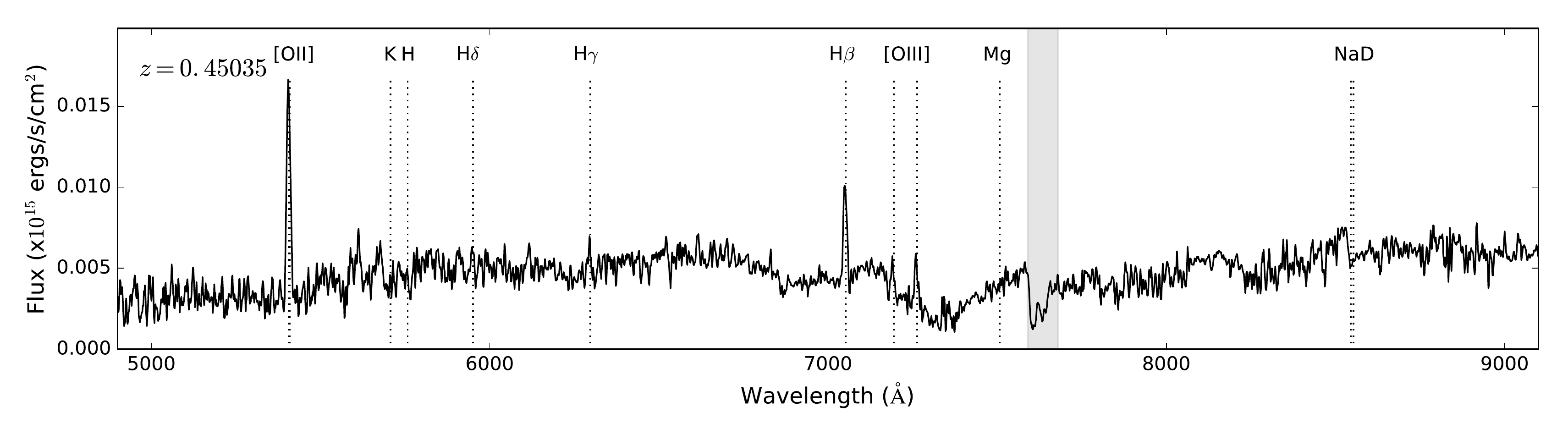}
\caption[]{Optical spectra of two of the galaxies along the line of sight to PKS\,1610-77, taken with the 8\,m Gemini-South telescope. (Top) Spectrum of Galaxy A at $z=0.45043$, showing strong NaD absorption. (Bottom) Spectrum of Galaxy B at $z=0.45035$, showing emission lines of H$\beta$, [OII] and [OIII]. }
\label{fig:1610-77_spec}
\end{figure*}

\subsubsection{The intervening galaxy group towards PKS\,1610-77}

The background QSO at $z=1.71$\ was studied in detail by \cite{courbin97}, who obtained an R-band optical image showing three galaxies within a few arcsec of the QSO position. A fourth galaxy was visible in their PSF-subtracted image. Figure \ref{fig:1610_gal2} shows a schematic view of the positions of the four galaxies found by \cite{courbin97}, relative to the radio-loud QSO targeted by ASKAP.  

\cite{courbin97} obtained detailed photometry of the region surrounding the radio-loud QSO, as well as optical spectra of the QSO and a nearby stellar object. 
They measured R-band magnitudes of 21.3, 21,3, 22,5 and 23.0 respectively for galaxies A, B, C and D, but were uncertain whether these galaxies were associated with PKS\,1610-77 itself or foreground objects along the line of sight. 
They also noted that their QSO spectrum appeared highly reddened in the optical, possibly by absorbing objects along the line of sight.  

\cite{courbin97} also noted the presence of a strong, unidentified absorption feature at 8,552\AA\ in their optical spectrum of PKS\,1610-77. We can now identify this feature as 5895.6\AA\ NaD absorption at a redshift of $z=0.4506$, i.e. the same redshift as the HI absorption line. NaD is a good tracer of cold neutral gas because of its low ionization potential \cite[e.g.][]{schwartz04}, so the identification of this optical line provides independent confirmation of the presence of cold neutral gas at the redshift of our ASKAP HI detection.  

 In 2018, we obtained optical spectra of the two brightest galaxies in the PKS\,1610-77 field (galaxies A and B) with the 8m Gemini-South telescope (see Figure  \ref{fig:1610-77_spec}). The redshift measured for both these galaxies ($z=0.4504$) is within 30\,km\,s$^{-1}$ of the ASKAP HI absorption redshift ($z=0.4503$), implying that the intervening HI gas is associated with this galaxy group. 
At z=0.45, the impact parameters of galaxies A, B, C and D are 10.0, 17.8, 24.2, and 5.9 kpc respectively. The separation between galaxies A and D is only 2.7\,arcsec, corresponding to a projected linear distance of $\sim16$\,kpc. 
With no k-correction applied, the approximate R-band absolute magnitudes of the four galaxies are $-20.7$ (A), $-20.7$ (B), $-19.5$ (C) and $-19.0$ (D).  

At this stage, it remains unclear which of the four galaxies in the group is the host galaxy of the HI gas seen in absorption. The spectrum of Galaxy A shows strong NaD absorption, which is absent from the spectrum of Galaxy B, but the spectrum of Galaxy B shows strong H$\beta$ emission characteristic of ongoing star formation. No spectrum is currently available for Galaxy D, which is the closest to the QSO line of sight, and further observations are needed to resolve this question.  

\subsubsection{The galaxy lens towards PKS\,1830-211}
This is a well-studied system, though dust extinction (and a density of foreground stars) within the Milky Way means that optical studies are extremely  difficult at the low Galactic latitude of PKS\,1830-211 ($b=-5.7$\,deg).

\cite{courbin02} and \cite{winn02} have identified the lensing galaxy as a face-on spiral, and 
\cite{koopmans05} showed that the HI absorption could be modelled by an almost face on gaseous disk with a constant rotation velocity and a radially dependent 21-cm optical depth. 

\cite{winn02} also estimated the lens magnification as a factor $\sim5.9$ for the NE component and $\sim3.9$ for the SW component of the lensed QSO. 
PKS\,1830-211 is one of the brightest radio sources in our sample. Even if we assume that all the radio emission in this source has been boosted by a factor of six, its unlensed counterpart would still have been bright enough to satisfy the selection criteria in \S2.2 of this paper. We therefore retain PKS\,1830-211 in our analysis despite the likelihood that its observed radio flux density has been increased by lensing. 

\subsubsection{The intervening system towards PKS\,0834-20 }
We have not yet identified a host galaxy for the HI absorption system at $z=0.591$ along the line of sight to PKS\,0834-20. A Pan-STARRS \citep{magnier13} image of the field shows faint red objects around 3.6\,arcsec west and 4.8\,arcsec south-east of the QSO. If these are galaxies at the same redshift at the HI line, the impact parameters would be roughly 25\,kpc and 32\,kpc respectively. It is currently unclear whether either of these objects is associated with the HI gas seen in absorption.   

\begin{figure}
\centering
\includegraphics[width=0.48\textwidth]{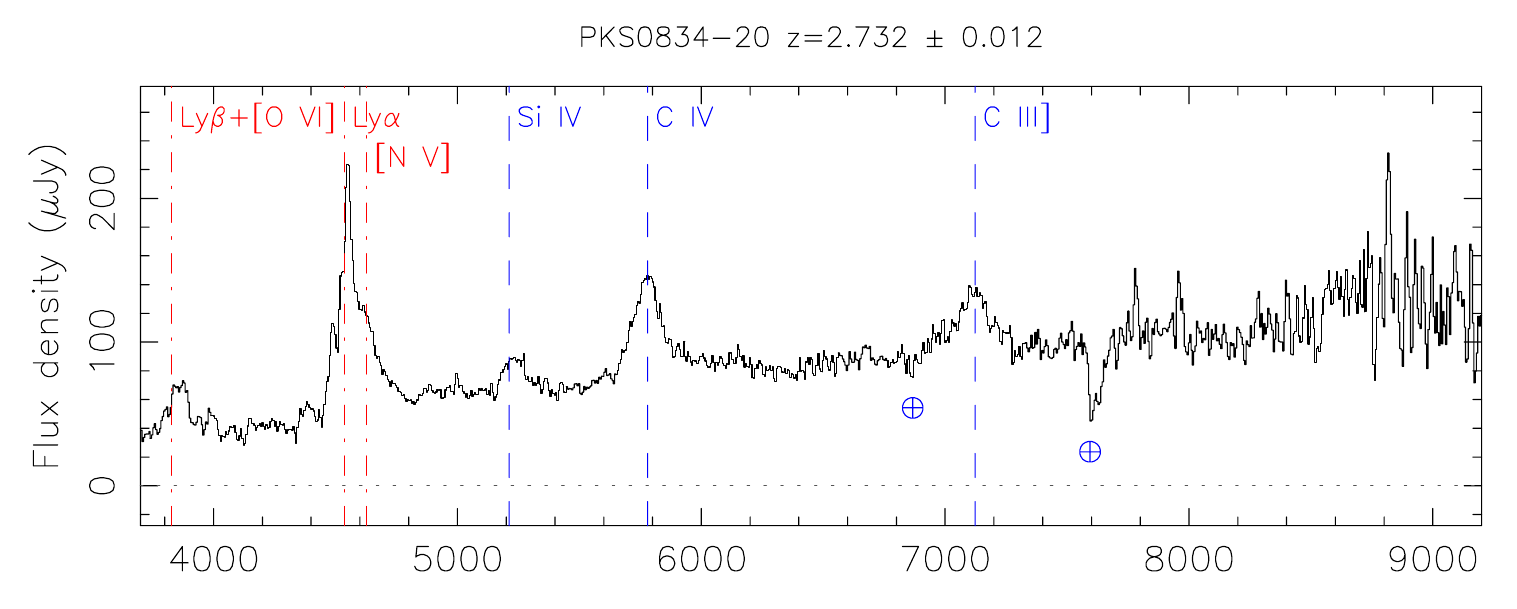}
\includegraphics[width=0.5\textwidth]{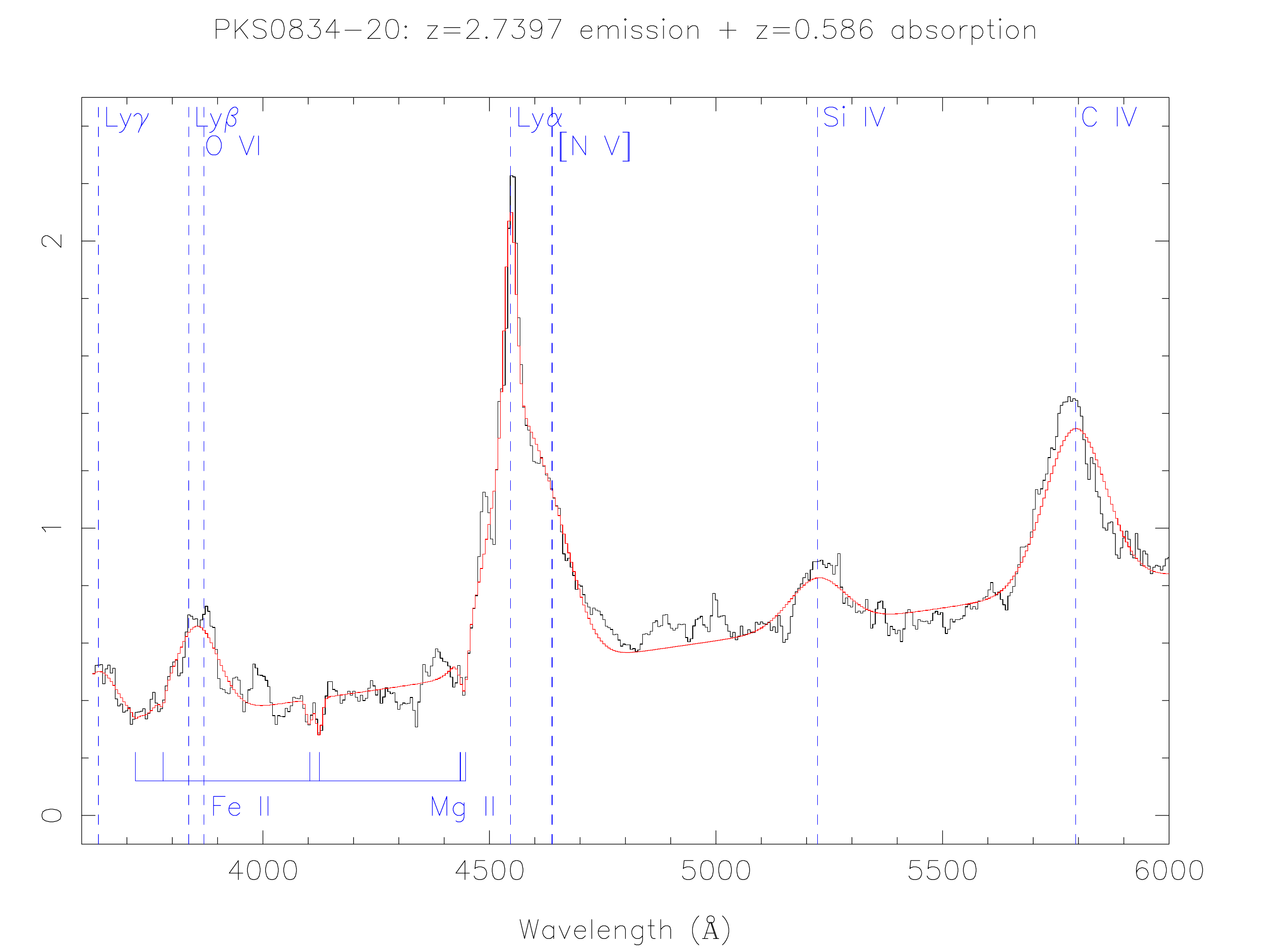}
\caption[]{(Top) Optical spectrum of the quasar PKS\,0834-20, taken with the 3.5m ESO NTT in March 2018, with the main spectral lines labelled. (Bottom) Close-up of the region blueward of Ly$\alpha$, showing metal absorption lines of FeII and MgII at $z=0.586$, close to the 21\,cm HI absorption redshift of $z=0.591$. }
\label{fig:0834_NTT}
\end{figure}

The only additional information we have at this stage comes from the optical spectrum of the background quasar. 
The top panel of Figure \ref{fig:0834_NTT} shows a spectrum of PKS\,0834-20 taken in 2018 with the ESO NTT, and the lower panel shows an expanded view of the blue region of this spectrum. Weak metal absorption lines of Fe\,II and Mg\,II are seen at a similar redshift to the intervening HI absorption at $z=0.591$. 

\ctable[  
notespar,
  star,
  caption={Potential host galaxies of the intervening HI absorption systems detected in this study},  
label={tab:hosts},
  ]{l l ll llll l ll cc} 
{\tnote[]{  } }
{\FL 
\multicolumn{1}{l}{Radio source}  & \multicolumn{1}{l}{$z {\small(\rm HI)}$} & \multicolumn{1}{l}{Potential} & \multicolumn{1}{l}{Mag.} & \multicolumn{1}{l}{Impact} & \multicolumn{1}{l}{Notes} \\
 & & \multicolumn{1}{l}{host} & & \multicolumn{1}{l}{parameter} \\
 \hline\hline
PKS\,0834-20 & 0.591 & Unknown  & .. & .. & No deep optical image available \\    
PKS\,1229-02 & 0.395 & Spiral galaxy & ..  & 2\,arcsec (11\,kpc) & See \cite{briggs85,kronberg92} \\
PKS\,1610-77 & 0.450 & Galaxy in group & 21.3 (R) & 1.7\,arcsec (10\,kpc) & If galaxy A, see \cite{courbin97}\\
PKS\,1830-211 & 0.886 & Spiral galaxy & .. &$<<1$\,arcsec ($<<7$\,kpc) & See \cite{winn02} 
\LL }

\subsubsection{The intervening galaxy towards PKS\,1229-02}
\cite{steidel94} obtained infrared I and K-band images of the PKS\,1229-02 field. After subtracting the QSO light, they found two faint objects likely to be foreground galaxies and tentatively identified the $z=0.395$ absorption system with a spiral galaxy at an impact parameter of $\sim7$\,kpc from the background QSO. 

This is consistent with the findings of \cite{kronberg92}, who carried out a detailed multi-frequency analysis of the rotation measure variations along the radio jet in PKS\,1229-02. They found that their results were consistent with the presence of an intervening spiral galaxy with an inclination angle of $\sim60^\circ$\ and an impact parameter of 2\,arcsec (around 10\,kpc at $z=0.395$). 
Hamanowicz et al. (2019, in preparation) have recently observed the PKS\,1229-02 field with the MUSE integral-field spectrograph on the ESO VLT. They measure a star-formation rate of $0.67\pm0.09$\,M$_\odot$\,yr$^{-1}$ for the intervening galaxy, and find that it has sub-solar metallicity (A. Hamanowicz, private communication). 

\section{Conclusions}\label{sec:conclusions}
In this pilot ASKAP study of the sightlines towards 53 bright, compact southern radio sources, we detected four intervening 21\,cm HI absorption lines at redshifts ranging from $z=0.395$ to $z=0.886$. Two of these (towards PKS\,1229-02 and PKS\,1830-211) are re-detections of lines from the published literature, and two (towards PKS\,0834-20 and PKS\,1610-77) are detected here for the first time. 

We used these detections to make a new estimate of the DLA number density at redshift $z\sim0.6$, $n(z)=0.19\substack{+0.15 \\ -0.09}$. This value lies above the general trend seen in optical and ultraviolet studies of QSO DLA systems \citep[e.g.][]{rao17,zafar13,noterdaeme12}, as can be seen from Figure \ref{fig:ndla}. 

From the small sample observed here, it appears that our  detected HI absorption lines arise mainly in the disks of spiral or late-type galaxies, with impact parameters typically less than 10--15\,kpc. At least one of the background radio-loud QSOs (PKS\,1610-77) is highly reddened (presumably by the intervening galaxy group identified by \cite{courbin97} and detected here in HI) and would probably not have been selected in an optical QSO survey. 

Our pilot results are encouraging for two reasons. 
Firstly, the redshift interval covered by a single ASKAP spectrum, $\Delta z\sim0.6$, is large enough to carry out spectroscopically untargeted 21\,cm searches for intervening HI absorption systems - a significant advance over earlier radio studies \citep[e.g.][]{lane98} where optical preselection (often based on the Mg\,II absorption line) was required. 
Secondly, the 700-1000\,MHz ASKAP band is free from terrestrial radio interference (in contrast to the case at most major radio observatories around the world), meaning that the full redshift interval can be searched for 21\,cm absorption lines. 

While this pilot survey has focused on observations of individual bright radio sources, the completed 36-antenna ASKAP telescope will have sufficient sensitivity to search for HI absorption simultaneously on sightlines to over 100 radio sources across a 30\,deg$^2$ field of view \citep{johnston08}. The full all-sky FLASH dataset will cover a total redshift path length $\Delta z \sim 50,000.$ This opens up the exciting possibility of detecting several hundred new 21\,cm absorption systems out to $z\sim1$, allowing us to improve our knowledge of the amount and physical state of neutral hydrogen in individual galaxies in the distant Universe. 

\section*{Data availability}
The data underlying this article will be shared on reasonable request to the corresponding author.

\section*{Acknowledgements}
We acknowledge the financial support of the Australian Research Council through grant CE170100013 (ASTRO 3D). 
The initial stages of this research were supported by the Australian Research Council Centre of Excellence for All Sky Astrophysics (CAASTRO), through grant CE110001020. 

The Australian SKA Pathfinder is part of the Australia Telescope National Facility which is managed by CSIRO. Operation of ASKAP is funded by the Australian Government with support from the National Collaborative Research Infrastructure Strategy. ASKAP uses the resources of the Pawsey Supercomputing Centre. Establishment of ASKAP, the Murchison Radio-astronomy Observatory and the Pawsey Supercomputing Centre are initiatives of the Australian Government, with support from the Government of Western Australia and the Science and Industry Endowment Fund. We acknowledge the Wajarri Yamatji people as the traditional owners of the Observatory site.

We gratefully acknowledge and thank the ASKAP commissioning team for the use of BETA and ASKAP-12 during the commissioning phase, which allowed us to carry out the observations described in this paper. 

Based on observations collected at the European Southern Observatory under ESO program 0100.A-0588(A). 

Based on observations obtained under program ID GS-2017B-Q-63 at the Gemini Observatory which is operated by the Association of Universities for Research in Astronomy, Inc., under a cooperative agreement with the NSF on behalf of the Gemini partnership: the National Science Foundation (United States), National Research Council (Canada), CONICYT (Chile), Ministerio de Ciencia, Tecnolog\'ia e Innovaci\'on Productiva (Argentina), Minist\'erio da Ci\^encia, Tecnologia e Inova\c c\~ao (Brazil), and Korea Astronomy and Space Science Institute (Republic of Korea).

This research has made use of the NASA/IPAC Extragalactic Database (NED), which is operated by the Jet Propulsion Laboratory, California Institute of Technology,
under contract with the National Aeronautics and Space Administration.

We thank C\'eline P\'eroux for helpful comments on an earlier draft of this paper, and Filippo Maccagni for carrying out one of the ESO NTT observing runs.  

EMS, VAM and JRA also thank the Munich Institute for Astro- and Particle Physics (MIAPP) for supporting our attendance at the 2019 MIAPP program on `Galaxy Evolution in a New Era of HI Surveys', which provided a stimulating venue for discussions and allowed us to complete this paper.  




\bibliographystyle{mnras}
\bibliography{references_intervening} 



\appendix
\section{Notes on individual sources} 

\noindent
{\bf PKS\,0903-57:} This source is at low Galactic latitude ($b=-7.0$\,deg). \cite{thompson90}\ quote a redshift of z=0.695$\pm$0.003, based on lines of redshifted MgII\,2798 seen at 4751\,\AA\ and [OII]\,3727 at 6306\,\AA, but the spectrum shown in their paper is noisy and the lines appear very weak. It also remains unclear whether the object observed is the correct optical ID for the radio source. We therefore regard the redshift of PKS\,0903-57 as uncertain at present. \\

\noindent
{\bf MRC\,1039-474:} 
The J2000 ICRF position listed by \cite{ma98} (10:41:42.940 -47:40:06.53) is offset by about 18\,arcsec from the VLBI position listed by \cite{titov13} (10:41:44.650 -47:40:00.06). We adopt the \cite{titov13} position, which is consistent with both the AT20G position listed by \cite{murphy10} and the optical position of the background QSO. \\

\noindent
{\bf PKS\,1127-14}
An intervening 21\,cm HI absorption line at $z=0.313$ was detected by \cite{lane98} and \cite{chengalur00}. This line is outside the redshift range covered by the ASKAP observations in Tables 1 and 2, but has recently been studied in detail by \cite{peroux19} and \cite{chen19}.   \\

\begin{figure}
\centering
\includegraphics[width=0.48\textwidth]{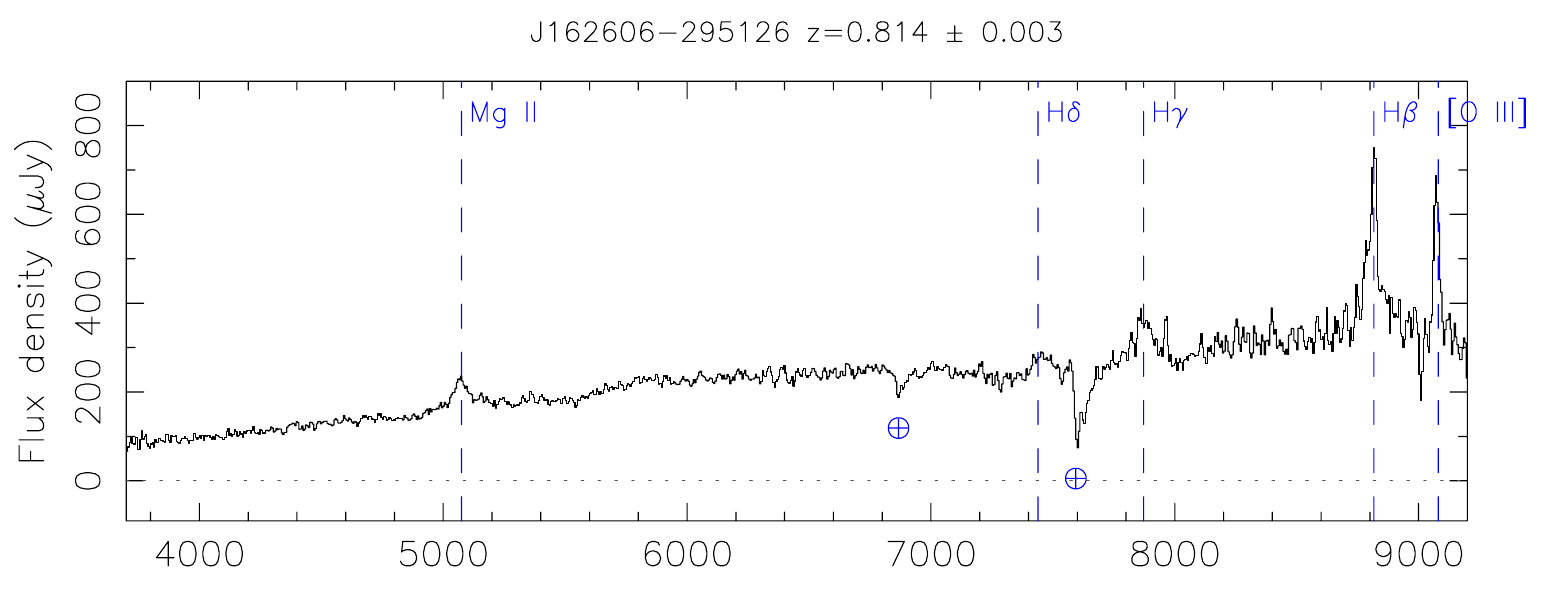}
\caption[]{Optical spectrum of the radio source  PKS\,1622-29, taken with the 3.5m ESO NTT in March 2018, with the main spectral lines labelled. }
\label{fig:1622_NTT}
\end{figure}

\noindent
{\bf PKS\,1622-29: } 
A redshift of z=0.815 is listed in the 1990 PKS catalogue, and is quoted in several published papers. As noted by \cite{jackson02} however, the initial redshift measurement  cannot be traced back to any published source. 
A new spectrum taken with the ESO NTT in 2018 and reproduced in Figure \ref{fig:1622_NTT}, gives a redshift of $z=0.814$, and so confirms the original literature redshift. 
\\

\noindent
{\bf PKS\,1740-571: } This source has a 21\,cm HI absorption line at $z=0.441$ identified as an associated absorption system by \cite{allison15}. 
\\

\noindent
{\bf MRC\,1759-396: } NED lists a possible redshift of z=0.293 \citep{liang03}, but we were unable to trace this back to its original source. Instead, we adopt the redshift of z=1.319 measured by \cite{shaw12} from an ESO NTT spectrum. 
\\

\noindent
{\bf PKS\,2123-463: } \cite{jackson02} note that the early redshift measurements by \cite{savage81}, based on objective-prism spectra, is not reliable and so the redshift of this object remains unknown. \cite{dammando12} estimate a photometric redshift of $z\sim1.46$ based on SED fitting. 
\\

\noindent{\bf PKS\,2223-05: } NED lists a $z=0.4842$ QSO DLA system found by \cite{lanzetta95} in IUE spectra. No published HST spectrum is available to confirm this detection, and \cite{turnshek02} note that the lack of MgII and FeII absorption lines at this redshift makes it unlikely that this $z=0.4842$ system is a genuine DLA. \cite{chengalur00} failed to detect 21\,cm absorption at $z=0.484$, and we also see no absorption at this redshift in our ASKAP spectrum.


\bsp	
\label{lastpage}
\end{document}